\documentclass[aps,twocolumn,amsmath,amssymb,floatfix,pra,reprint,showpacs,footinbib,superscriptaddress,longbibliography]{revtex4-1}
\usepackage{epsfig}
\usepackage{times}
\usepackage{amsmath}
\usepackage{amsfonts}
\usepackage{amssymb}
\usepackage[usenames]{color}
\usepackage{graphicx}

\newcommand{\ex}[1]{\langle #1 \rangle}

\newcommand{\eq}[1]{Eq.~(\ref{#1})}
\newcommand{\fig}[1]{Fig.~\ref{#1}}

\newcommand{\ket}[1]{\ensuremath{\vert#1\rangle}}

\newcommand{\beq}{\begin{equation}}
\newcommand{\eeq}{\end{equation}}
\newcommand{\beqa}{\begin{eqnarray}}
\newcommand{\eeqa}{\end{eqnarray}}

\newcommand{\appref}[1]{\mbox{Appendix~\ref{#1}}}

\newcommand{\NTT}{NTT Basic Research Laboratories, NTT Corporation, 3-1
Morinosato-Wakamiya, Atsugi, Kanagawa, 243-0198, Japan.}

\newcommand{\riken}{CEMS, RIKEN, Wako-shi, Saitama 351-0198, Japan}

\newcommand{\mich}{Department of Physics, University of Michigan, Ann Arbor, Michigan 48109-1040, USA}
\begin{document}
\title{Superradiance with
\textcolor{black}{an ensemble of superconducting flux qubits}
}
\author{Neill Lambert}		\email{nwlambert@riken.jp}\affiliation{\riken}
 \author{Yuichiro Matsuzaki}			\email{matsuzaki.yuichiro@lab.ntt.co.jp} \affiliation{\NTT}
 \author{Kosuke Kakuyanagi}			\affiliation{\NTT}
  \author{Natsuko Ishida}		\affiliation{\riken}
 \author{Shiro Saito} 	\affiliation{\NTT}
 \author{Franco Nori} 	\affiliation{\riken}	\affiliation{\mich}

\begin{abstract}

Superconducting flux qubits are a promising candidate for realizing
 quantum information processing and quantum simulations. Such devices behave like
 artificial atoms, with the advantage that one can easily tune the ``atoms'' internal properties. Here, by harnessing this flexibility,
 we propose a technique to minimize the inhomogeneous broadening of a
 large ensemble of flux qubits by tuning only the external flux.  In
 addition, as an example of many-body physics
 in such an ensemble, we show how to observe superradiance, and its quadratic scaling with ensemble size, using a tailored
 microwave control pulse that takes advantage of the inhomogeneous broadening itself to excite only a sub-ensemble of the qubits. Our scheme opens up {an} approach to using
 superconducting circuits to explore the properties of quantum many-body systems.
\end{abstract}
\maketitle

\section{Introduction}
Superconducting flux qubits (FQ) are a unique quantum technology which allow for a high degree of controllability
\cite{You2011,ClarkeWilhelm01a,Buluta2011}. With such devices high-fidelity gate
operations have already been  implemented
\cite{bylander2011noise} and quantum non-demolition measurements have
been realized using Josephson bifurcation
amplifiers. 
Moreover, since superconducting FQs behave as controllable artificial atoms, it
is possible to design circuits
\textcolor{black}{ to reach regimes typically inaccessible with real atoms \cite{yoshihara2016superconducting,chen2016multi,forn2016ultrastrong}. }

\begin{figure}[ht]
\includegraphics[width=\columnwidth]{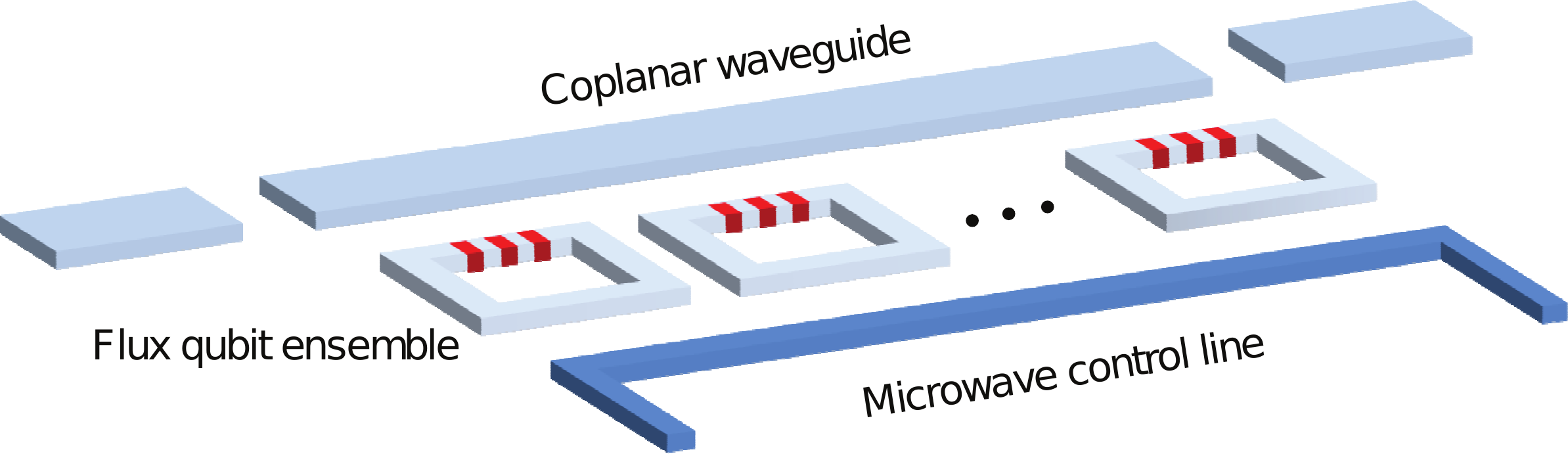}
\caption{ (Color online) Schematic of a potential flux qubit ensemble system. We estimate upto 4300 FQs can be coupled with a microwave
   cavity. \textcolor{black}{One may characterize this system by measuring the transmission through
the cavity.
}
 }
\label{device}%
\end{figure}

As well as featuring high-controllability, flux qubits are attractive because it is possible to fabricate an array of
FQs on the same chip~\cite{kakuyanagi4300}.
Coupling such an array of
{\em many} superconducting FQs to a common cavity (see  Fig. \ref{device} for a schematic) is important both for a range of quantum
information processing tasks and for the study of quantum
many-body physics~\cite{Buluta108,rmp2014}, \textcolor{black}{like quantum}
phase
transitions \cite{hepp1973superradiant,
wang1973phase,emary2003chaos,lambert2009quantum,lambert2004entanglement}. In addition, an array of superconducting FQs could be
used as a quantum metamaterial to control the propagation of
microwaves
\cite{PhysRevB.77.144507,soukoulis2011past,zheludev2012metamaterials,macha2014implementation}.
\textcolor{black}{Such a device also allows for the possibility of generating multi-particle entanglement between the FQs via
the cavity, with the potential to employ this entanglement to improve the
sensitivity of measurements
\cite{kitagawa1993squeezed,Ma2011,bennett2013phonon,tanaka2015proposed}.}

\textcolor{black}{One obstacle to such applications with an
ensemble of FQs is the inhomogeneity of the FQ energies. }
 In the context of strong coupling to a cavity, this can be overcome to some degree by using the
superradiance principle
\cite{imamouglu2009cavity,wesenberg2009quantum,macha2014implementation}; if $N$ qubits
are collectively coupled with a microwave cavity, the coupling strength
is enhanced by $\sqrt{N}$,
\textcolor{black}{as long as the collective coupling strength is larger than the
inhomogeneous width }
\cite{rmp2013hybrids,schuster2010high,wu2010storage,kubo2010strong,amsuss2011cavity,kubo2011hybrid,zhu2011coherent,kubo2012storage,
kubo2012electron,marcos2010coupling,twamley2010superconducting,matsuzaki2012enhanced,saito2013towards,julsgaard2013quantum,
diniz2011strongly,zhudark2014}.
{Recently,} {by using this principle, coupling between 4300 superconducting flux
qubits and a microwave resonator has been demonstrated \cite{kakuyanagi4300}.
In this experiment, spectroscopic measurements were performed by detecting the transmitted
photon intensity of the resonator, and  a large
dispersive shift of $250$ MHz has been observed. This already
indicates a collective behavior involving thousands of
 {FQ}s.  }

{In this paper,} we discuss how the intrinsic inhomogeneity can be
reduced \textcolor{black}{by a globally applied external field,
}
\textcolor{black}{an effect which we will show to be a direct consequence of the correlation
between the tunneling energy and persistent current in FQs.
}
In addition, we show how, as one
of the potential applications of this device,
one can observe
superradiant emission from such an ensemble via the microwave cavity.
Superradiance is the \textcolor{black}{fascinating}
phenomena whereby an ensemble of atoms
interacting with a common cavity or environment emits photons
in a fast, collective, superradiant burst, due to correlations between atomic decay
events.
\textcolor{black}{For this type of superradiance, the loss rate of the cavity needs to
be larger than the collective coupling of the ensemble with the cavity
mode, while the collective coupling strength should be much larger than
the inhomogeneous width of the FQ ensemble.
}
 The observation of superradiance provides a direct signal
 \textcolor{black}{of the collective coupling between the ensemble and the
 common field.}

To date superradiance  has been observed in various many-particle systems
\cite{skribanowitz1973observation,gross1976observation,raimond1982collective,scheibner2007superradiance,rohlsberger2010collective}.
In addition, there are some experimental demonstrations of superradiance with only small ensembles of engineered quantum systems
\cite{devoe1996observation,eschner2001light,filipp2011multimode,van2013photon,mlynek2014observation}.
Typically the observation of this superradiant burst requires the
careful preparation of all the atoms in their excited states,
and the subsequent observation of the time-dependent photonic intensity (though steady-state driven superradiance can also occur under the right
conditions \cite{Meiser2010}).
In the latter half of this article we show theoretically that we can prepare
the ensemble of {FQ}s with a common drive, and see not only the
typical large intensity superradiance emission pulse,
but also the $N^2$ scaling of that pulse, without local control of each qubit.

This paper is organized as follows.
Firstly, we review the recent experimental spectroscopic
measurements
to explain the standard properties of the system.
Secondly, we introduce a scheme to suppress
the inhomogeneous broadening of the FQs, which is crucial to observe
superradiance and other many-body properties of such a system. Finally, we present numerical results showing how
collective driving of the ensemble can selectively excite the ensemble, allowing us to directly observe the $N^2$ superradiant emission.

\section{Spectroscopic measurements}
The first experimental test one could make to validate a coupling between the
ensemble and the cavity is to look for vacuum Rabi splitting or
frequency shift in spectroscopic measurements.
In a recent experiment, spectroscopic measurements of
the microwave resonator coupled with 4300 {FQ}s \cite{kakuyanagi4300} showed a large dispersive frequency shift, in the spectrum of the cavity, of the order of
250 MHz.
\textcolor{black}{Although similar signals of collective behavior have been observed in many
other systems \cite{rohlsberger2010collective,keaveney2012cooperative,roof2016observation}, for a system composed of
a large FQ ensemble and a microwave resonator,}
this is the first strong signature of a large collective
coupling \cite{kakuyanagi4300}.
There, the coupling strength between a single {FQ} and the
resonator was estimated to be around $15$ MHz, and the inhomogeneous width of
the {FQ} frequency was between $2$ and $3$ GHz.
Interestingly, even if there is an inhomogeneous width
of a few GHz, a clear dispersive frequency shift can be observed,
because the collective coupling strength
($\sqrt{N}\overline{g}\simeq 1$ GHz) is comparable with the
inhomogeneous width.
It is worth mentioning that, in principle, one can increase this coupling strength
by using a Josephson junction as a coupler
\cite{yoshihara2016superconducting}, and so one could achieve the ultra-strong coupling regime \cite{Chen16USC, Forn16USC}
with this system
where $\sqrt{N}g$ is both much larger than the inhomogeneous width and of the order of the flux qubit and cavity energies themselves.

\section{Suppression of the inhomogeneous broadening}

To observe superradiance in such an ensemble, the collective coupling strength $\sqrt{N}g$
should be larger than the variance of the frequency distribution of
the FQs. \textcolor{black}{Moreover, to invert the FQs using a global microwave control, the Rabi frequency of the FQs should also be larger than
the inhomogeneous width, as we will describe later.
However, from the direct parameters estimated in \cite{kakuyanagi4300},
it is difficult to satisfy such conditions.
}

  \begin{figure}[ht]
\includegraphics[width=1.01\linewidth]{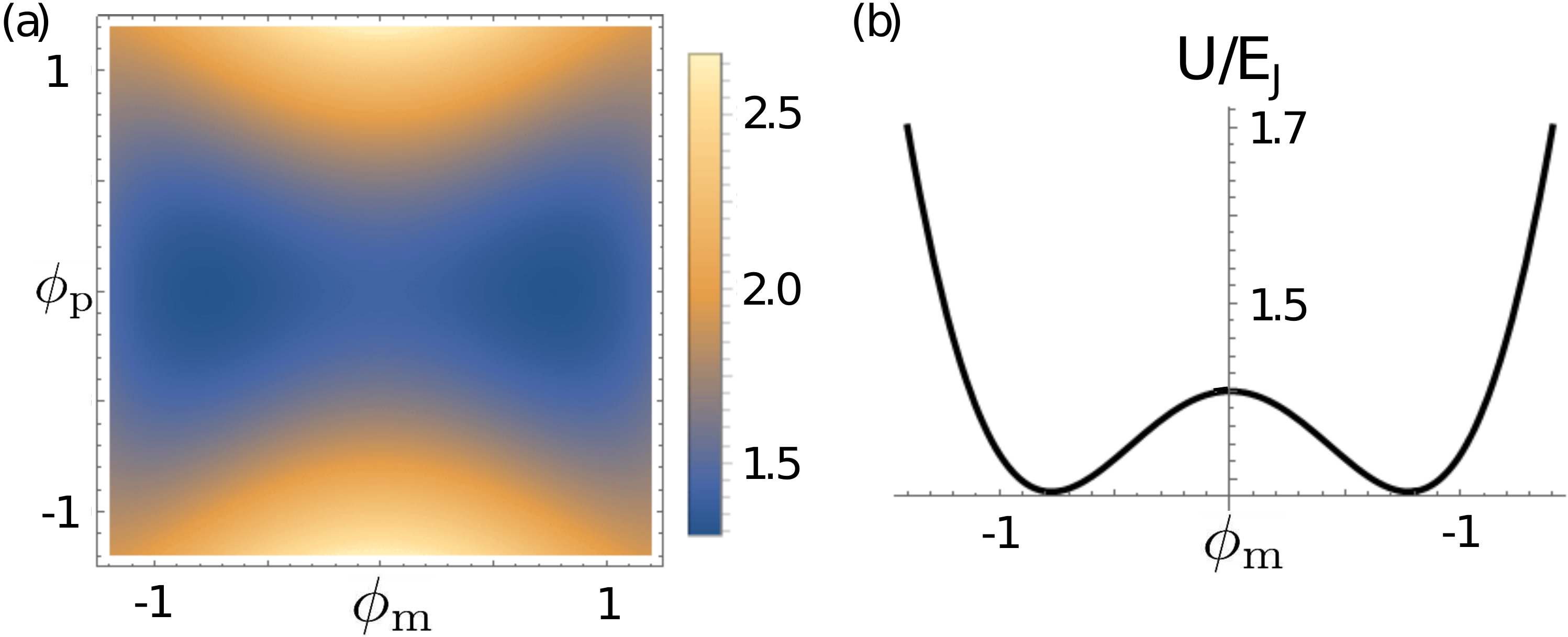}
\caption{ (Color online) The potential of the flux qubit.
   We set $\alpha
   =0.7$ and $f=0.5$. There are two minima separated by an energy
   barrier.
   (a) The density plot of the potential. (b) A plot of the potential
   against $\phi _m$ for $\phi _p=0$.
 }
\label{potential}%
\end{figure}

\textcolor{black}{To solve these problems, we propose here an approach to}
suppress the inhomogeneous broadening
 of the {FQ}s by applying an external magnetic flux.
 The inhomogeneous broadening of the FQ energies comes from the
non-uniform size of the Josephson junctions,  which are very sensitive
to small changes in fabrication conditions.
We have investigated how
the non-uniform Josephson junctions affect the relevant parameters of the {FQ}s, and have found that
the variation of the size of the Josephson junctions
induces a {\em correlated} distribution between the persistent current and
tunneling energy of the {FQ}s in the ensemble. Interestingly, due to this correlation,
the inhomogeneous width of the frequencies of the {FQ}s has a strong dependence on the applied magnetic flux, and so there
exists the possibility of choosing an
optimal applied magnetic flux to suppress this
broadening. We predict this property could be useful to design more uniform ensembles of quantum devices, thus allowing
us to observe interesting quantum many-body phenomena, such as superradiance.

To investigate how the non-uniform Josephson junctions affect the
frequency distributions of a {FQ}, we consider
the Lagrangian of a {FQ} with three Josephson junctions
\begin{eqnarray}
 L&=&T-U \\
 U&=&\sum\limits_{j=1}^{3}
\frac{\Phi_{0}}{2\pi}I_{\text{C}}^{j}\left[1-\text{cos}(\phi_{j}]\right) \\
 T&=&\sum\limits_{j=1}^{3} \frac{1}{2}C_{j}\left(\frac{\Phi_{0}}{2\pi}\right)^{2}\dot{\phi_{j}}^{2}
\end{eqnarray}
where $U$ is the potential energy, $T$ is the
kinetic energy, $\phi_{j}$ $(j=1,2,3)$ is
the phase difference between the junctions, $C_{j}$ is the
 the Josephson junction capacitance, $I_{\text{C}}^{j}$ is
 the critical current,  $\Phi_{\text{ext}}$ is
 the external magnetic flux, and
 $\Phi_{0}=\hbar/2e$ is the magnetic flux quantum.
The phases $\phi _j$ $(j=1,2,3)$ are bounded by a condition of $\phi_{1}-\phi_{2}+\phi_{3}=2\pi f$ with
$f=\Phi{\text{ext}}/\Phi_{0}$.
$C_j$ and $I_{\text{C}}^{j}$ have a linear dependence on the size of
 the junction.
 \textcolor{black}{Here, the potential is given by
$U/E_J=2+\alpha -\text{cos}(\phi_{\text{p}}+\phi_{\text{m}})-
   \text{cos}(\phi_{\text{p}} -\phi_{\text{m}}) -\alpha \text{cos}(2\pi f -2\phi_{\text{m}})$
where we set $I^1_{\text{C}}=I^2_{\text{C}}=I_{\text{C}}$,
 $I^3_{\text{C}}=\alpha I_{\text{C}}$,
 $\phi_{\text{p}}=(\phi_{1}+\phi_{2})/2$, and
 $\phi_{\text{m}}=(\phi_{1}-\phi_{2})/2$. If we set $\phi
 _{\text{p}}=0$ and $f=0.5$, we have $\frac{dU}{df}=2E_J\sin \phi _m
 (1-2\alpha \cos \phi _m)$, and so the potential shows minima for
 $\pm \phi _{\text{m}}^*$, where $\cos \phi ^*_{\text{m}}=1/(2\alpha)
 $. We plot this potential in Fig. \ref{potential}.}
 By solving the Lagrangian, we can calculate the
 tunneling energy and persistent current
 \cite{orlando1999superconducting}.
We set $E_J/E_c=75$ for our simulations,  where
$E^{(j)}_J=\frac{\Phi_{0}}{2\pi}I_{\text{C}}^{j}$
($E_c=e^2/2C_j$) is the characteristic scale of the Josephson (electric) energy.

    \begin{figure}[ht]
\begin{center}
\includegraphics[width=0.9\columnwidth]{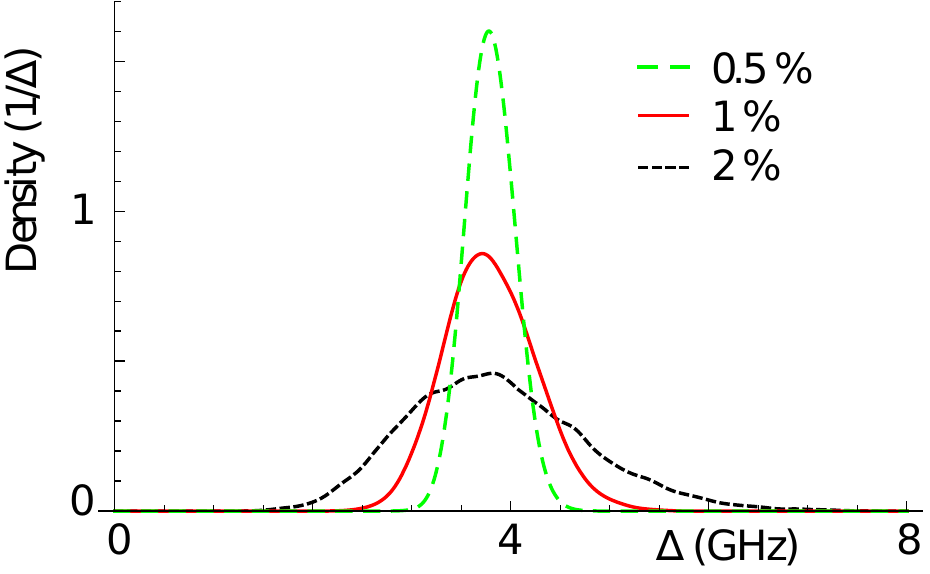}
\caption{
 (Color online) The probability density of the tunneling energies of the flux qubits
 when the size of the Josephson
 junctions are non-uniform.
 There are three Josephson
 junctions in the superconducting circuit, and the size of one Josephson junction
 is designed to be smaller than the size of the other two
 junctions.
 We assume a Gaussian distribution for normalized areas of the smaller
  junction (two larger
  junctions)
  where we have the mean value of $\bar{\alpha }=0.7$
  ($\bar{\beta }_k=1$ for $k=1,2$) and the standard
  deviation of $\sigma _{\text{S}}$ ($\sigma ^{(k)}_{\text{L}}$ for
  $k=1,2$). We set the parameters as $\sigma _{\text{S}}/\bar{\alpha }=\sigma
 ^{(1)}_{\text{L}}/\bar{\beta }=\sigma ^{(2)}_{\text{L}}/\bar{\beta
 }=0.5\ \%, 1\ \%,2\ \%$ respectively, and obtain
 the values of $\Delta _j$ ($j=1,2,\cdots ,N$) from numerical simulations.
To plot the density of the tunneling energy, we use a kernel density
 estimator $\sum_{j=1}^{N}K(\frac{\Delta -\Delta _j}{h})$, where we set
 $K(x)=\frac{1}{\sqrt{2\pi }}\exp{\left(-\frac{1}{2}x^2\right)}$, $N=10000$, and $h=0.1$ GHz.
 }
 \label{delta}
\end{center}

\end{figure}

 Usually, the size of one of the three junctions is designed to be $\alpha $ times
 smaller than the other two junctions \cite{orlando1999superconducting}.
  However, with current technology it is difficult to fabricate
homogenous junctions, and this results in a random distribution of the
tunneling energy and the persistent current.
We assume a Gaussian distribution for the normalized areas of the smaller
  junction (two larger
  junctions),
  where we have the mean value of $\bar{\alpha }$
  ($\bar{\beta }_k$ for $k=1,2$) and the standard
  deviation of $\sigma _{\text{S}}$ ($\sigma ^{(k)}_{\text{L}}$ for
  $k=1,2$).

  Firstly, in  Fig. \ref{delta} we plot the distribution of the tunneling energies of the {FQ}. This confirms that the non-uniform
  Josephson junctions affect the random distribution of the tunneling
  energy.
  As expected, as we increase the width of the distribution of the Josephson junction
  size, the
  width of the tunneling energy distribution also increases.

\begin{figure}[ht]
\begin{center}
\includegraphics[scale=0.3]{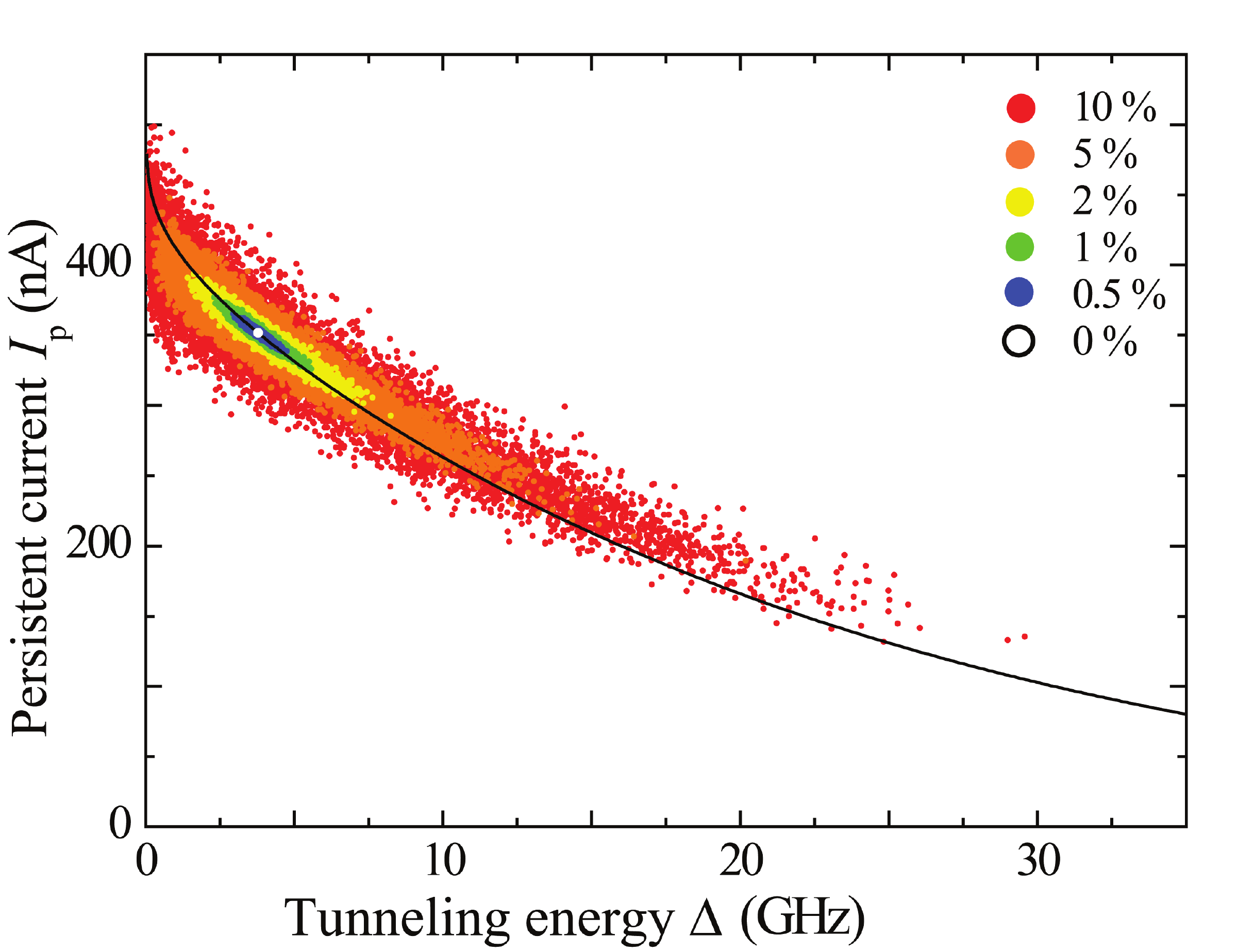}
\caption{(Color online) The persistent currents and
  tunneling energies of FQs with random-size Josephson junctions.
 We set the same parameters as in Fig. \ref{delta}. There is a clear
 correlation between the tunneling energy $\Delta$ and persistent current $I_p$.
 }
 \label{correlate}
\end{center}
\end{figure}

  Secondly, we plot the distribution of the persistent current and
  tunneling energy given by the non-uniform Josephson junctions in
  Fig. \ref{correlate}.
  We randomly generate the values of the Josephson junction size, and calculate
  the resulting tunneling energy and persistent current. This result clearly show a correlation between the tunneling
  energy and persistent current where a {FQ} with a higher
  tunneling energy tends to have a lower persistent current.
  \textcolor{black}{We can qualitatively explain this correlation as
  follows.
 As we increase the value of
 $\alpha $, the potential gradient $\frac{dU}{df}\simeq 2\pi E_J
  \left[1-\frac{1}{(2\alpha )^2}\right]^{1/2}$  becomes larger for $\phi _{\text{p}}\simeq 0$, $\phi
  _{\text{m}} \simeq \phi _{\text{m}}^*$, and $f\simeq 0.5$. A larger
  potential gradient makes the energy
  of the {FQ} more sensitive to
 the change in the applied magnetic flux, which corresponds to a
 higher persistent current. On the other hand, as we increase the value
  of $\alpha$, the tunneling barrier $E_{\text{t}}=U(\phi
  _{\text{m}}=0)-U(\phi _{\text{m}}=\phi _{\text{m}}^*)=E_J(-2+2\alpha
  +\frac{1}{2\alpha })$ becomes larger for $\phi _{\text{p}}\simeq 0$, $\phi
  _{\text{m}} \simeq \phi _{\text{m}}^*$, $\alpha \simeq 0.7$, and
  $f\simeq 0.5$. The larger tunneling barrier  suppresses the
  tunneling energy of the {FQ}. Therefore, if the persistent current
  becomes larger, the tunneling energy is expected to be smaller, which is
  consistent with our numerical simulations. 
Moreover, it is worth mentioning that a similar model was used to reproduce the
  experimental results in \cite{kakuyanagi4300} where spectroscopy of a microwave
  resonator coupled to 4300 {FQs} was performed and good agreement between
  numerical and experimental results was observed \cite{kakuyanagi4300}.
   In that experiment,  the standard
  deviation of the Josephson junction size is around a few percent,
  which corresponds to the yellow region in Fig.  \ref{correlate}.}

  \begin{figure}[ht]
\begin{center}
\includegraphics[scale=0.3]{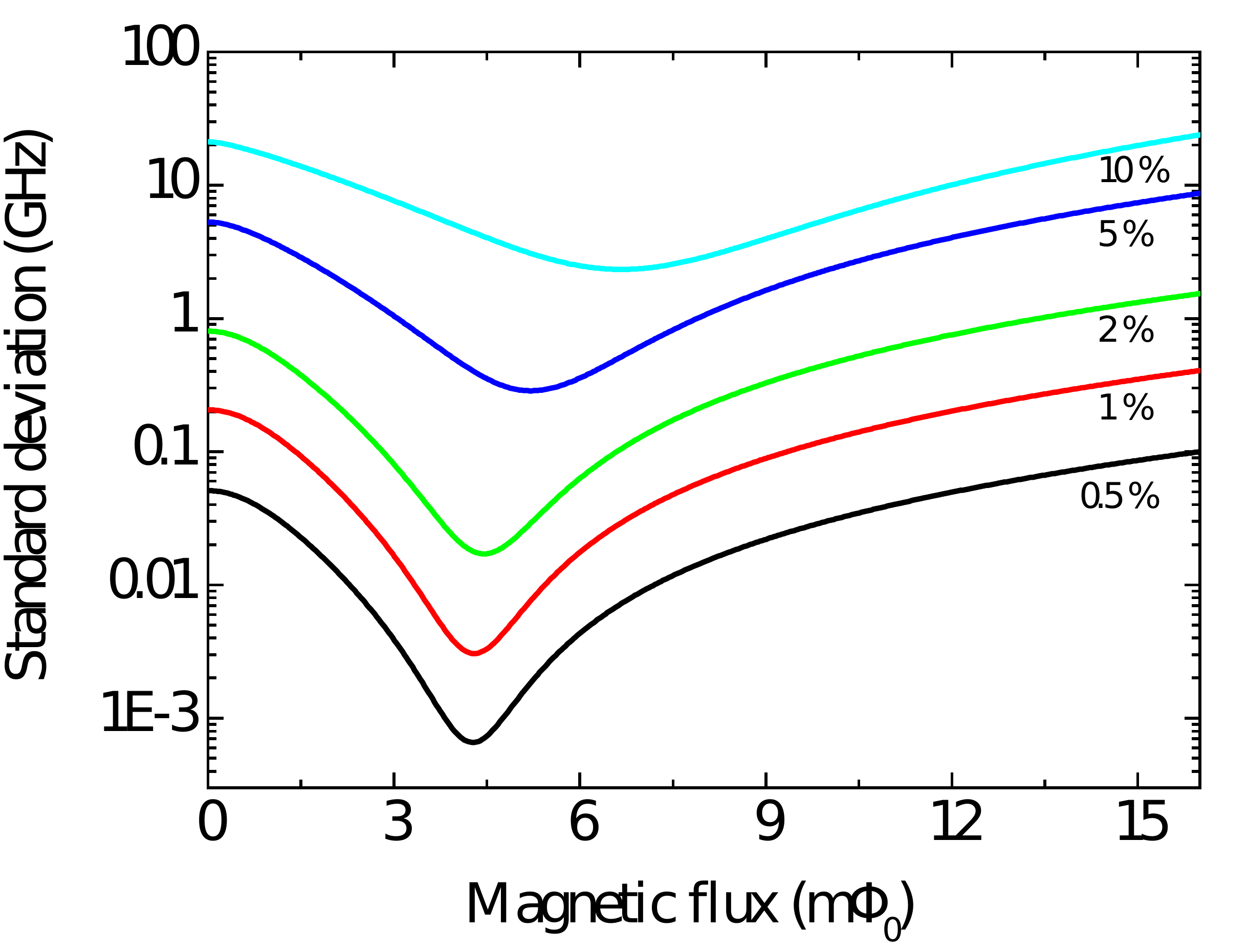}
\caption{(Color online) The standard deviation of the distribution of the flux qubit frequencies
 versus the applied magnetic flux.
  We set the same parameters as in Fig. \ref{delta}.
 The standard deviation strongly depends on the applied magnetic flux.
 }
 \label{suppress}
\end{center}
\end{figure}

  \begin{figure}[ht]
\begin{center}
\includegraphics[scale=0.3]{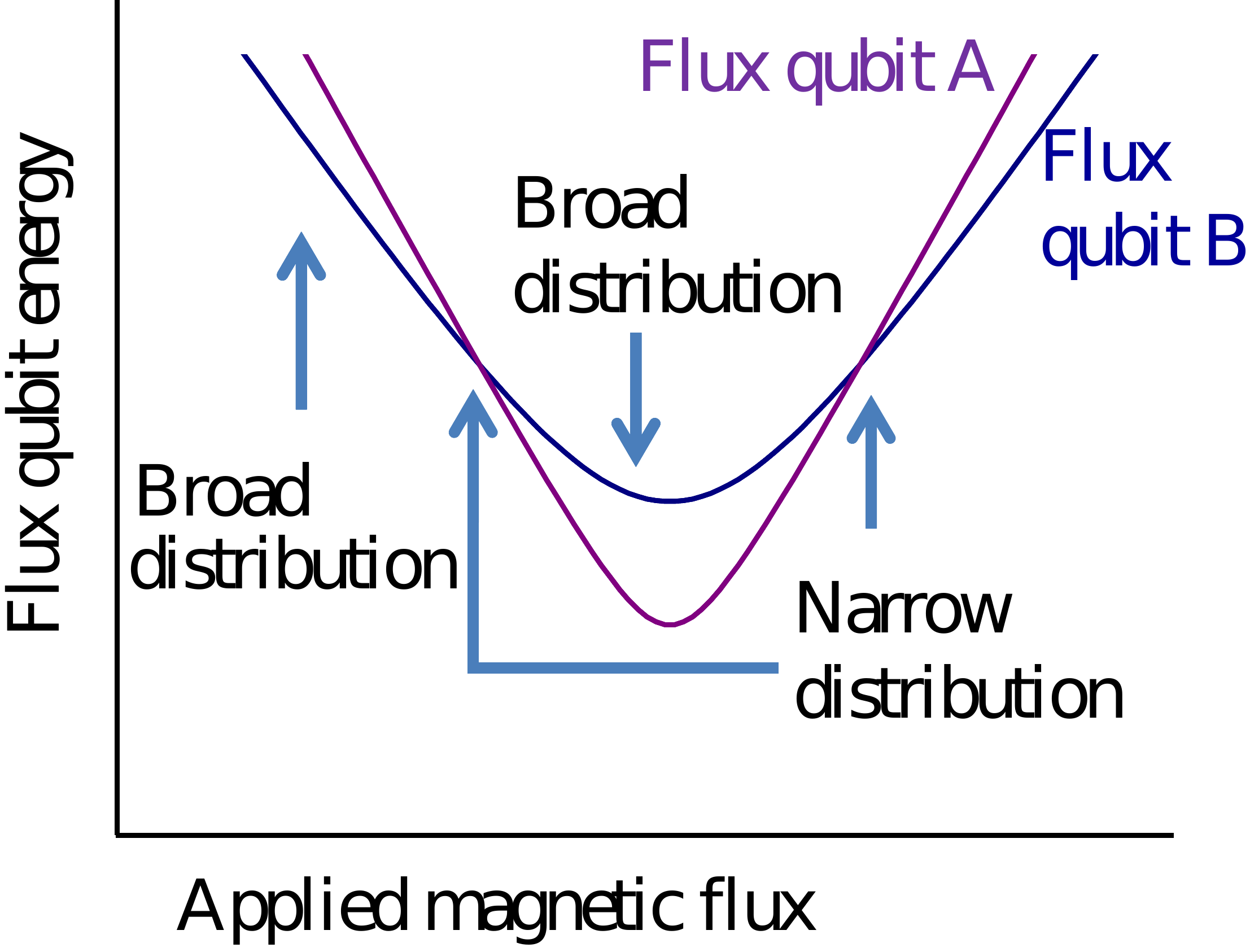}
\caption{
Energies of two flux qubits (A and B), with different size junctions, as a function of an applied magnetic field.
 The flux qubit energy is represented by $\omega
  _j=\sqrt{|\epsilon _j|^2 +|\Delta ^{(\text{t})}_j|^2}$ for $\epsilon _j=2I_j(\Phi _{\text{ext}}-\frac{1}{2}\Phi
_0)$ where $\Phi _{\text{ext}}$ denotes the applied magnetic flux.
 Flux qubit A has a smaller (larger) tunneling energy (persistent
 current) than B.
 In this case, we can make the frequency of the qubits the same
 by applying an appropriate amount of the applied magnetic flux.
 }
 \label{schematic-supress}
\end{center}
\end{figure}

  Thirdly, in Fig. \ref{suppress} we plot the standard deviation of the {FQ} frequency
  distribution against an applied magnetic field.
  Interestingly, these results show that the standard deviation of
  the frequency distribution strongly depends on the applied magnetic flux;
  there exists an optimal point where the standard deviation of the flux
  qubit frequency becomes minimum.
  The width of the distribution becomes one or two orders
  of magnitude smaller at the optimal point than elsewhere.
  This can be understood as a consequence of the correlation between the tunneling energy
  and the persistent current, as shown in
  Fig. \ref{schematic-supress}.

To illustrate this idea, let us consider a pair of
  flux qubits with different junction sizes.
  The {FQ} energy is given by $\omega
  _j=\sqrt{|\epsilon _j|^2 +|\Delta ^{(\text{t})}_j|^2}$, for $\epsilon _j=2I_j(\Phi _{\text{ext}}-\frac{1}{2}\Phi
_0)$ $(j=1,2)$, and we can assume $\Delta ^{\text{(t)}}_1> \Delta^{\text{(t)}}_2 $ without loss of
  generality.
 \textcolor{black}{ Interestingly, when $I_1 < I_2$, which is the expected statistical relationship given $\Delta_1 > \Delta_2$, we can show that there exists an optimal flux such that
$\omega _1=\omega_2$ is satisfied. }
  So we can balance the two flux qubit energies just by applying a global magnetic flux.
This means that, even if we have several qubits with
  different-size Josephson junctions, 
   if
  there is a correlation such that
a smaller  persistent current $I_j$
    tends to increase
    the tunneling energy $\Delta
    ^{(\text{t})}_j $, we can
 make the
  frequency of these qubits similar by tuning an external magnetic flux, as shown in Fig. \ref{correlate}.

\section{Superradiance}

\begin{figure}[!t]
\includegraphics[width=1.\columnwidth]{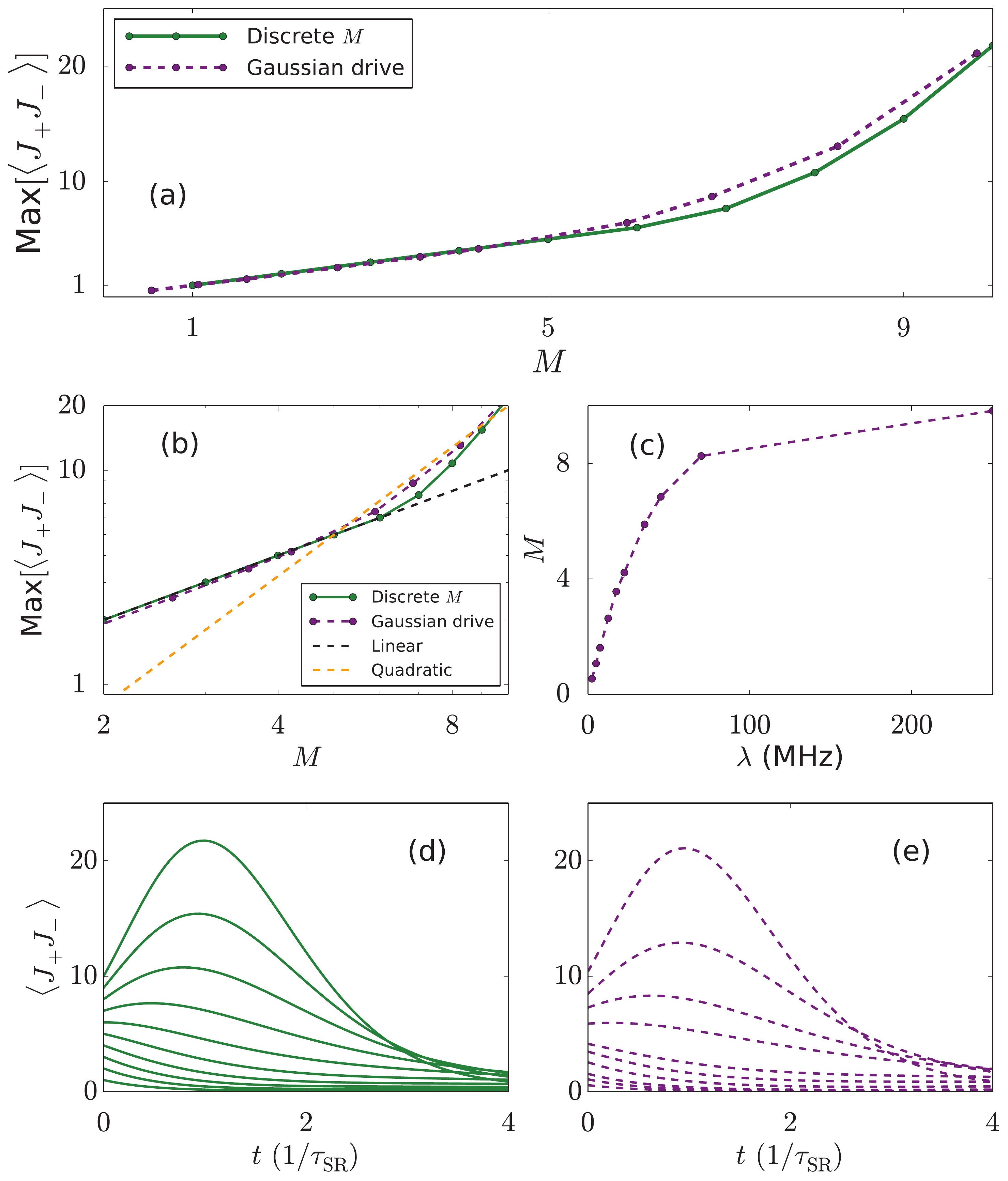}
\caption{ (a) Maximum (over time) emitted intensity versus number of initially-excited qubits $M$. For discrete
 $M$ (green solid curve) we artificially prepare a subset $M$ of the total ensemble of  $N$ qubits in their
 excited states.
 In the other case (purple dashed line), at $t=0$ we prepare  all qubits in their ground state
 and then evolve with the Hamiltonian  $H_{\mathrm{drive}}(t)$ switched
 on,
 with a Gaussian
 function envelope $\lambda(t)$, as described in the main text.
 We then switch off the driving and allow the system to evolve under the
 influence of \eq{ME}, and record the maximum emitted intensity
 over a time period exceeding the expected superradiant pulse
 duration. We do this for a range of $\lambda_{{\rm max}}$,
 which induce an effective number $M$ of qubits to
 become excited.
 For the other parameters we set $\bar{\omega}_j=\omega_c$,
 $\delta \omega_j=25$ MHz, $g=50$ MHz, $\kappa=400$ MHz, $N=10$, so as to give a
 value for
 $\tilde{\alpha} > 1$
 as $M$ becomes greater than $4$.
 Panel (b) shows the logarithmic intensity, which changes from
 linear to quadratic behavior as $M$ passes this $M=4$ threshold (the gray
 dotted line is an artificial
 linear comparison curve, while the orange dotted line is an artificial quadratic comparison curve, to show this change clearly).
 Panel (c) shows the explicit $\lambda_{\mathrm{max}}$ values used in the Gaussian drive, and the associated number of excited spins $M$ in the ensemble after the drive has been applied.
 Panel (d) shows the explicit time-dependent curves of intensity for different
 values of $M$, increasing from the bottom up, starting with $M=1$ to $10$, from which  the green dotted-dashed line in figure (a) is
 extracted. The change from normal
 to superradiant emission around $M=4$ is clear.  Similarly, panel (e) shows the same curves for the driven state preparation example.  Note that all curves are averaged over a large set of randomly-generated ensemble energies.}\label{SR}
\end{figure}
To illustrate how such an ensemble with a reduced inhomogeneous width can lead to observable collective effects, we numerically simulate~\cite{qutip1,qutip2} a small
ensemble with an explicit inhomogeneity. We also show how this residual inhomogeneity can be used as a tool to aid initial-state preparation.
We explicitly model $N=10$ {FQ}s, with
 inhomogeneous normally-distributed energies $\omega_j$ with
 mean value $\bar{\omega}_j$ and variance $\delta \omega_j$.  These qubits are
 coupled to a single common microwave cavity of frequency $\omega_c$
 with a common homogenous coupling strength $g$.   The general Hamiltonian for such a system reads,
\beq
H =  \sum_{j=1}^N\frac{\omega_j}{2} \sigma_z^{(j)} + \omega_c a^{\dagger}a + g\left(J_- a^{\dagger}+J_+a\right),
\label{HD}
\eeq
where $J_+=\sum_j \sigma_+^{(j)}$, $J_-=\sum_j \sigma_-^{(j)}$, and we have set $\hbar=1$ for simplicity. In
 general we assume that the cavity decay, with rate $\kappa$, is given
 by a Lindblad superoperator $\kappa \mathcal{D}[a]$, where $D[a]=2a \rho a^{\dagger} - a^{\dagger}a\rho  - \rho a^{\dagger}a$.

To begin with, we eliminate the cavity~\cite{Temnov05,Delanty11}, assuming the
bad-cavity limit: $\kappa \gg {\delta \omega_j}, g^2 N/\kappa$
(superradiance is also possible in the dispersive
good-cavity limit, see \appref{App}). In this bad-cavity case the
equation of motion is reduced to the following form
\beq
H_{\mathrm{AE}}=  \sum_{j=1}^N\frac{\omega_j-\bar{\omega}_j}{2}\sigma_z^{(j)} +
(\omega_c-\bar{\omega}_j) \frac{g^2}{\Gamma^2} J_+J_-
                                \label{aehamiltonian}
\eeq
where $\Gamma=\kappa +i(\omega_c - \bar{\omega_j})$. There also arises a new loss term, $\mathcal{S}{[\rho]}=\kappa\frac{g^2}{\Gamma^2} \mathcal{D}[J_-]\rho$.
It is this term that induces the superradiance phenomena, and we expect to observe such superradiance when ${\delta \omega_j } \ll g^2 N/\kappa$.

Even though the cavity is eliminated, one can estimate the intensity of the radiation emitted from the qubits from the squared atomic polarization \cite{Temnov05},
\beq
I(t) = \frac{2g^2}{\kappa} \omega_c \ex{J_+(t)J_-(t)}.
\eeq
Typically the intensity grows with time, reaches a maximum at the peak superradiance time
$\tau_{{\rm sr}}=\kappa/g^2 N$  and then decays. The uccessful observation of
this pulse
requires that the coherence time of the qubits is longer than the
expected peak superradiance time. Assuming dephasing is dominated by
the inhomogeneity of the energies of the FQs, we can assess the
visibility of superradiance via the parameter
$\tilde{\alpha}=T_2^*/\tau_{{\rm sr}}=Ng^2/\kappa \delta
\omega_j$
{where $T_2^*$ is the inhomogeneous dephasing time}.


In addition to the qubits being inhomogeneous, the direct control of
individual qubits is challenging. However, we can consider 
collective ways
in which to prepare spin-polarized states, which we can be used to
observe superradiance.  In particular, by strongly driving the cavity, or using another common control line, as per  Fig. \ref{device},
we can induce a time-dependent collective control term, such that the dynamics of the qubits can be written as,
\begin{eqnarray}
H_{\mathrm{drive}}&=&\sum_{j=1}^N\frac{\omega_j}{2}\sigma_z^{(j)}+\lambda(t)\cos (\omega_d t)\sum_j \sigma_x^{(j)}, \\
H_{\mathrm{drive}}'&\approx& \sum_{j=1}^N\frac{\Delta_j' }{2}\sigma_z^{(j)}+\frac{\lambda(t)}{2} \sum_j \sigma_x^{(j)},\label{HD2}
\end{eqnarray}
where in the second equation we moved to a frame rotating at the drive frequency, such that $\Delta_j' =
\omega_j - \omega_d$, and made a rotating-wave approximation. Later we will choose the drive to be resonant with the average value of the qubit energies
$\omega_d =\bar{\omega}_j$. If we consider just a single qubit,
initially in its ground state,  we know  that if we apply a drive of strength
$\lambda$ for a period $T_{\pi}= \pi/ \lambda$ we will find that the spin has a probability of being in its excited state:
\beq
P_{\mathrm{exc}}=\frac{\lambda^2}{\Delta_j'^2 +\lambda^2},
\eeq

Extending this notion to $N$ spins we expect that we will have an
effective excited number of spins
$M_{\mathrm{eff}}=\sum_j \frac{\lambda^2}{\Delta_j'^2 +\lambda^2}$.
Thus, simply changing the magnitude of $\lambda$ enables
us to effectively control the number of spins contributing to the
superradiance emission (up to the limit of validity of the rotating
wave approximation). In addition, one can also control the shape of the envelope of the drive,
$\lambda(t)$. While $P_{\mathrm{exc}}$ and $M_{\mathrm{eff}}$
only apply for a step-function envelope, they provide a useful estimate.
In practise we found that a Gaussian function for $\lambda(t)$ worked best in preparing the desired initial state, and
thus only show that example here.
In principle one can also use more sophisticated techniques from
quantum control theory to prepare the desired state
\cite{mintert2,mintert1,Zhang16}.

\textcolor{black}{Importantly, when we need to excite most of the
qubit ensemble, the drive, or Rabi frequency, $\lambda $ should be as
large or larger than the inhomogeneous width. Although it is possible to achieve a Rabi
frequency of a few GHz \cite{yoshihara2014flux} for a single FQ, it is
not straightforward to realize such a strong driving condition for a large ensemble.
For this reason, it is crucial to decrease the
width of the inhomogeneous broadening of the FQs, by, for example, applying a magnetic
flux, as described earlier. This will allow us to both excite the ensemble with moderate values of $\lambda$, and observe superradiance
with accesible values of $g$. }

To obtain numerical results we solve the master equation for all $N$
qubits explicitly by generating a random ensemble of energies,
preparing the qubit ensemble in the common ground state (without
interaction with the cavity)
$\psi(0)=\ket{0}_1 \bigotimes \ket{0}_2 \bigotimes \ldots
\bigotimes\ket{0}_N$, and then ``switch on" the driving term
$H_{\mathrm{drive}}'(t)$
for a period $\tau$ such that $\int_0^\tau \lambda(t)\approx \pi$.  We
assume that during this driving period the cavity and qubit
ensemble are far off-resonance. In other words, the ensemble evolves under the free evolution of the ensemble Hamiltonian and the drive, given by
$H_{\mathrm{drive}}'(t)$ in \eq{HD2}, without influence from the cavity. In principle this implies we also require that the period $\tau$ is
shorter than the relaxation time of the qubits.

After this evolution, we record the effective number of excited qubits $M=\ex{\sum_j \sigma_z^{(j)}}$, switch off the drive, and allow the system to evolve
under both $H_{\mathrm{AE}}$ and the superradiant
loss term $\mathcal{S}[\rho]=\kappa\frac{g^2}{\Gamma^2} \mathcal{D}[J_-]\rho$, as determined by the master equation
\beq \label{ME}
\dot{\rho}=-\frac{i}{\hbar}[H_{\mathrm{AE}},\rho]+\mathcal{S}[\rho]
\eeq
for a time interval much longer than $\tau_{{\rm sr}}$ (recalling
$\tau_{{\rm sr}}=\kappa/g^2 M$, where $M$ are the number of qubits excited by the drive).
For this period of evolution we record the cavity emission intensity by
calculating $I(t)$,
and from this measurement record the maximum (over time) acquired value Max[$\ex{J_+J_-}$]. Under perfect superradiance Max[$\ex{J_+J_-}$] should scale as $M^2$.

We repeat this procedure as a function of the driving strength $\lambda$,
and plot the recorded maximum intensity Max[$\ex{J_+J_-}$] as
a function of $M$, the effective number of qubits initially  excited by the drive. Figure \ref{SR} shows this for a Gaussian drive shape
$\lambda(t)=\lambda_{\rm max}  \exp
\left[-\left(\frac{t-b}{\sigma}\right)^2\right]$ with $\sigma =
\sqrt{\pi}/\lambda_{\rm max}$ and $b=4\sigma\sqrt{2\ln{2}}$.
This is
compared to the test case where the actual number of initially excited qubits is
enforced ``by hand'', which we refer to as the ``discrete $M$" case.  We now see that the drive prepares a subset of the qubits in their
excited states,
thus altering the resultant photonic emission intensity. This allows us to directly observe the quadratic scaling of that intensity as a
function
of the number of qubits contributing to the collective decay.  For the parameters chosen here, we see the onset
of superradiance when $M$ becomes greater than about four (see
caption of \fig{SR}).

 In practice, as the number of FQs increases, one can still see superradiance
for much larger values of the inhomogeneity, or smaller couplings, than we show here.
For example, from the simulations described above,
 we can extrapolate the behavior
 of a device composed of 4300 FQs coupled with the microwave cavity.
 Due to the form of the loss term $\mathcal{S}{[\rho]}=\frac{g^2}{\kappa
 } \mathcal{D}[J_-]\rho$, for $\omega _{\rm {c}}=\overline{\omega }_j$, we
 should have a similar behavior for the emitted intensity from the
 cavity, as long as the value of $M g^2/\kappa$ is the same.
 Thus, if we fabricate a device with $g=5$ MHz, $\delta \omega _j=25$ MHz, $\kappa =1.72$ GHz, and $N=4300$, and excite the full ensemble,
 so that $M=N$,
 the value of $M g^2/\kappa$ coincides with that used in our
 numerical simulation with $10$ excited qubits; and so we should be able to observe the quadratic
 scaling of the intensity for this case as well.
 This means that one can see superradiance from
 4300 FQs
 even for coupling
 strengths as small as $5$ MHz.

\section{Conclusions}

We have shown that, even though large ensembles of FQs suffer from intrinsic fabrication-induced inhomogeneities, this can be minimized by tuning the
ensemble FQs properties with an external flux.  This opens up the possibility of observing collective many-body effects, a simple example of which we give in terms of superradiant emission
into a microwave cavity.  We expect that such large ensembles will enable the investigation of a range of interesting physics in the future, including criticality~\cite{hepp1973superradiant,
wang1973phase,emary2003chaos,lambert2009quantum,lambert2004entanglement},
macroscopic coherence \cite{Knee2016,lambert16}, and
\textcolor{black}{ spin squeezing
 \cite{kitagawa1993squeezed,Ma2011,bennett2013phonon,tanaka2015proposed}.}

\acknowledgments

\textcolor{black}{This work was supported by JSPS
KAKENHI Grant 15K17732, JSPS KAKENHI Grant
No.25220601, the Commissioned Research No. 158 of
NICT, and MEXT KAKENHI Grant Number  15H05870.  FN and NL acknowledge support from the Sir John Templeton Foundation.
FN acknowledges support from the RIKEN iTHES Project, the MURI Center for Dynamic Magneto-Optics via the AFOSR award number FA9550-14-1-0040, the IMPACT program of JST, CREST, and a Grant-in-Aid for Scientific Research (A).
N. L. and Y. M. contributed equally to this work.
}
\appendix

\section{Dispersive superradiance model. \label{App}}

One can also obtain collective superradiant decay due to interaction with a common cavity by moving to a dispersive
coupling regime \cite{bennett2013phonon}, where the cavity and qubits are
off-resonance,
without necessarily demanding that the cavity losses be large. Starting again with
\eq{HD} one can apply the transformation $e^{R} H_D e^{-R}$,
where $R= \frac{g}{\chi}(J_- a^{\dagger}- J_+a)$, $\chi=\omega_c - \bar{\omega_j}$, and keeping terms to order $(g/\chi)^2$ find that,
\beq
H_{\mathrm{disp}}=  \sum_{j=1}^N(\frac{1}{2}\omega_j+\beta a^{\dagger}a) \sigma_z^{(j)} + \frac{\beta}{2} J_+J_-
\eeq
where $\beta=2g^2/\chi$ and again a new loss term arises,
\beq
\mathcal{S}_{\mathrm{disp}}=\kappa\frac{g^2}{\chi^2} \mathcal{D}[J_-]\rho
\eeq
One expects in this case that superradiance will occur when $g^2 N
\kappa/\chi^2 \gg \delta\omega_j$, giving an equivalent parameter to assess
the visibility $ \alpha_D=g^2 N \kappa/(\chi^2\delta\!\omega_j)$.
However, this regime is valid for $(g/\chi)^2 \ll 1$, which implies
$N \kappa / \delta\!\omega_j \gg (g/\chi)^2$.  As with the adiabatic elimination
case, the spin squeezing term $J_+J_-$ does not affect the superradiance
dynamics significantly.


\begin{thebibliography}{64}%
\makeatletter
\providecommand \@ifxundefined [1]{%
 \@ifx{#1\undefined}
}%
\providecommand \@ifnum [1]{%
 \ifnum #1\expandafter \@firstoftwo
 \else \expandafter \@secondoftwo
 \fi
}%
\providecommand \@ifx [1]{%
 \ifx #1\expandafter \@firstoftwo
 \else \expandafter \@secondoftwo
 \fi
}%
\providecommand \natexlab [1]{#1}%
\providecommand \enquote  [1]{``#1''}%
\providecommand \bibnamefont  [1]{#1}%
\providecommand \bibfnamefont [1]{#1}%
\providecommand \citenamefont [1]{#1}%
\providecommand \href@noop [0]{\@secondoftwo}%
\providecommand \href [0]{\begingroup \@sanitize@url \@href}%
\providecommand \@href[1]{\@@startlink{#1}\@@href}%
\providecommand \@@href[1]{\endgroup#1\@@endlink}%
\providecommand \@sanitize@url [0]{\catcode `\\12\catcode `\$12\catcode
  `\&12\catcode `\#12\catcode `\^12\catcode `\_12\catcode `\%12\relax}%
\providecommand \@@startlink[1]{}%
\providecommand \@@endlink[0]{}%
\providecommand \url  [0]{\begingroup\@sanitize@url \@url }%
\providecommand \@url [1]{\endgroup\@href {#1}{\urlprefix }}%
\providecommand \urlprefix  [0]{URL }%
\providecommand \Eprint [0]{\href }%
\providecommand \doibase [0]{http://dx.doi.org/}%
\providecommand \selectlanguage [0]{\@gobble}%
\providecommand \bibinfo  [0]{\@secondoftwo}%
\providecommand \bibfield  [0]{\@secondoftwo}%
\providecommand \translation [1]{[#1]}%
\providecommand \BibitemOpen [0]{}%
\providecommand \bibitemStop [0]{}%
\providecommand \bibitemNoStop [0]{.\EOS\space}%
\providecommand \EOS [0]{\spacefactor3000\relax}%
\providecommand \BibitemShut  [1]{\csname bibitem#1\endcsname}%
\let\auto@bib@innerbib\@empty
\bibitem [{\citenamefont {You}\ and\ \citenamefont {Nori}(2011)}]{You2011}%
  \BibitemOpen
  \bibfield  {author} {\bibinfo {author} {\bibfnamefont {J.~Q.}\ \bibnamefont
  {You}}\ and\ \bibinfo {author} {\bibfnamefont {F.}~\bibnamefont {Nori}},\
  }\bibfield  {title} {\enquote {\bibinfo {title} {{Atomic physics and quantum
  optics using superconducting circuits}},}\ }\href {\doibase
  10.1038/nature10122} {\bibfield  {journal} {\bibinfo  {journal} {Nature}\
  }\textbf {\bibinfo {volume} {474}},\ \bibinfo {pages} {589} (\bibinfo {year}
  {2011})},\ \Eprint {http://arxiv.org/abs/1202.1923} {arXiv:1202.1923}
  \BibitemShut {NoStop}%
\bibitem [{\citenamefont {Clarke}\ and\ \citenamefont
  {Wilhelm}(2007)}]{ClarkeWilhelm01a}%
  \BibitemOpen
  \bibfield  {author} {\bibinfo {author} {\bibfnamefont {J.}~\bibnamefont
  {Clarke}}\ and\ \bibinfo {author} {\bibfnamefont {F.~K.}\ \bibnamefont
  {Wilhelm}},\ }\bibfield  {title} {\enquote {\bibinfo {title} {Superconducting
  quantum bits},}\ }\href@noop {} {\bibfield  {journal} {\bibinfo  {journal}
  {Nature}\ }\textbf {\bibinfo {volume} {453}},\ \bibinfo {pages} {1031}
  (\bibinfo {year} {2007})}\BibitemShut {NoStop}%
\bibitem [{\citenamefont {Buluta}\ \emph {et~al.}(2011)\citenamefont {Buluta},
  \citenamefont {Ashhab},\ and\ \citenamefont {Nori}}]{Buluta2011}%
  \BibitemOpen
  \bibfield  {author} {\bibinfo {author} {\bibfnamefont {I.}~\bibnamefont
  {Buluta}}, \bibinfo {author} {\bibfnamefont {S.}~\bibnamefont {Ashhab}}, \
  and\ \bibinfo {author} {\bibfnamefont {F.}~\bibnamefont {Nori}},\ }\bibfield
  {title} {\enquote {\bibinfo {title} {Natural and artificial atoms for quantum
  computation},}\ }\href@noop {} {\bibfield  {journal} {\bibinfo  {journal}
  {Rep. Prog. Phys.}\ }\textbf {\bibinfo {volume} {74}},\ \bibinfo {pages}
  {104401} (\bibinfo {year} {2011})}\BibitemShut {NoStop}%
\bibitem [{\citenamefont {Bylander}\ \emph {et~al.}(2011)\citenamefont
  {Bylander}, \citenamefont {Gustavsson}, \citenamefont {Yan}, \citenamefont
  {Yoshihara}, \citenamefont {Harrabi}, \citenamefont {Fitch}, \citenamefont
  {Cory}, \citenamefont {Nakamura}, \citenamefont {Tsai},\ and\ \citenamefont
  {Oliver}}]{bylander2011noise}%
  \BibitemOpen
  \bibfield  {author} {\bibinfo {author} {\bibfnamefont {J.}~\bibnamefont
  {Bylander}}, \bibinfo {author} {\bibfnamefont {S.}~\bibnamefont
  {Gustavsson}}, \bibinfo {author} {\bibfnamefont {F.}~\bibnamefont {Yan}},
  \bibinfo {author} {\bibfnamefont {F.}~\bibnamefont {Yoshihara}}, \bibinfo
  {author} {\bibfnamefont {K.}~\bibnamefont {Harrabi}}, \bibinfo {author}
  {\bibfnamefont {G.}~\bibnamefont {Fitch}}, \bibinfo {author} {\bibfnamefont
  {D.~G.}\ \bibnamefont {Cory}}, \bibinfo {author} {\bibfnamefont
  {Y.}~\bibnamefont {Nakamura}}, \bibinfo {author} {\bibfnamefont {J.~S.}\
  \bibnamefont {Tsai}}, \ and\ \bibinfo {author} {\bibfnamefont {W.~D.}\
  \bibnamefont {Oliver}},\ }\bibfield  {title} {\enquote {\bibinfo {title}
  {Noise spectroscopy through dynamical decoupling with a superconducting flux
  qubit},}\ }\href@noop {} {\bibfield  {journal} {\bibinfo  {journal} {Nature
  Physics}\ }\textbf {\bibinfo {volume} {7}},\ \bibinfo {pages} {565--570}
  (\bibinfo {year} {2011})}\BibitemShut {NoStop}%
\bibitem [{\citenamefont {Yoshihara}\ \emph {et~al.}(2016)\citenamefont
  {Yoshihara}, \citenamefont {Fuse}, \citenamefont {Ashhab}, \citenamefont
  {Kakuyanagi}, \citenamefont {Saito},\ and\ \citenamefont
  {Semba}}]{yoshihara2016superconducting}%
  \BibitemOpen
  \bibfield  {author} {\bibinfo {author} {\bibfnamefont {F.}~\bibnamefont
  {Yoshihara}}, \bibinfo {author} {\bibfnamefont {T.}~\bibnamefont {Fuse}},
  \bibinfo {author} {\bibfnamefont {S.}~\bibnamefont {Ashhab}}, \bibinfo
  {author} {\bibfnamefont {K.}~\bibnamefont {Kakuyanagi}}, \bibinfo {author}
  {\bibfnamefont {S.}~\bibnamefont {Saito}}, \ and\ \bibinfo {author}
  {\bibfnamefont {K.}~\bibnamefont {Semba}},\ }\bibfield  {title} {\enquote
  {\bibinfo {title} {Superconducting qubit-oscillator circuit beyond the
  ultrastrong-coupling regime},}\ }\href@noop {} {\bibfield  {journal}
  {\bibinfo  {journal} {arXiv preprint arXiv:1602.00415}\ } (\bibinfo {year}
  {2016})}\BibitemShut {NoStop}%
\bibitem [{\citenamefont {Chen}\ \emph {et~al.}(2016)\citenamefont {Chen},
  \citenamefont {Wang}, \citenamefont {Li}, \citenamefont {Tian}, \citenamefont
  {Qiu}, \citenamefont {Inomata}, \citenamefont {Yoshihara}, \citenamefont
  {Han}, \citenamefont {Nori}, \citenamefont {Tsai} \emph
  {et~al.}}]{chen2016multi}%
  \BibitemOpen
  \bibfield  {author} {\bibinfo {author} {\bibfnamefont {Zhen}\ \bibnamefont
  {Chen}}, \bibinfo {author} {\bibfnamefont {Yimin}\ \bibnamefont {Wang}},
  \bibinfo {author} {\bibfnamefont {Tiefu}\ \bibnamefont {Li}}, \bibinfo
  {author} {\bibfnamefont {Lin}\ \bibnamefont {Tian}}, \bibinfo {author}
  {\bibfnamefont {Yueyin}\ \bibnamefont {Qiu}}, \bibinfo {author}
  {\bibfnamefont {Kunihiro}\ \bibnamefont {Inomata}}, \bibinfo {author}
  {\bibfnamefont {Fumiki}\ \bibnamefont {Yoshihara}}, \bibinfo {author}
  {\bibfnamefont {Siyuan}\ \bibnamefont {Han}}, \bibinfo {author}
  {\bibfnamefont {Franco}\ \bibnamefont {Nori}}, \bibinfo {author}
  {\bibfnamefont {JS}~\bibnamefont {Tsai}},  \emph {et~al.},\ }\bibfield
  {title} {\enquote {\bibinfo {title} {Multi-photon sideband transitions in an
  ultrastrongly-coupled circuit quantum electrodynamics system},}\ }\href@noop
  {} {\bibfield  {journal} {\bibinfo  {journal} {arXiv preprint
  arXiv:1602.01584}\ } (\bibinfo {year} {2016})}\BibitemShut {NoStop}%
\bibitem [{\citenamefont {Forn-Diaz}\ \emph {et~al.}(2016)\citenamefont
  {Forn-Diaz}, \citenamefont {Garcia-Ripoll}, \citenamefont {Peropadre},
  \citenamefont {Yurtalan}, \citenamefont {Orgiazzi}, \citenamefont
  {Belyansky}, \citenamefont {Wilson},\ and\ \citenamefont
  {Lupascu}}]{forn2016ultrastrong}%
  \BibitemOpen
  \bibfield  {author} {\bibinfo {author} {\bibfnamefont {P}~\bibnamefont
  {Forn-Diaz}}, \bibinfo {author} {\bibfnamefont {JJ}~\bibnamefont
  {Garcia-Ripoll}}, \bibinfo {author} {\bibfnamefont {B}~\bibnamefont
  {Peropadre}}, \bibinfo {author} {\bibfnamefont {MA}~\bibnamefont {Yurtalan}},
  \bibinfo {author} {\bibfnamefont {J-L}\ \bibnamefont {Orgiazzi}}, \bibinfo
  {author} {\bibfnamefont {R}~\bibnamefont {Belyansky}}, \bibinfo {author}
  {\bibfnamefont {CM}~\bibnamefont {Wilson}}, \ and\ \bibinfo {author}
  {\bibfnamefont {A}~\bibnamefont {Lupascu}},\ }\bibfield  {title} {\enquote
  {\bibinfo {title} {Ultrastrong coupling of a single artificial atom to an
  electromagnetic continuum},}\ }\href@noop {} {\bibfield  {journal} {\bibinfo
  {journal} {arXiv preprint arXiv:1602.00416}\ } (\bibinfo {year}
  {2016})}\BibitemShut {NoStop}%
\bibitem [{\citenamefont {Kakuyanagi}\ \emph {et~al.}(2016)\citenamefont
  {Kakuyanagi}, \citenamefont {Matsuzaki}, \citenamefont {Deprez},
  \citenamefont {Toida}, \citenamefont {Semba}, \citenamefont {Yamaguchi},
  \citenamefont {Munro},\ and\ \citenamefont {Saito}}]{kakuyanagi4300}%
  \BibitemOpen
  \bibfield  {author} {\bibinfo {author} {\bibfnamefont {K.}~\bibnamefont
  {Kakuyanagi}}, \bibinfo {author} {\bibfnamefont {Y.}~\bibnamefont
  {Matsuzaki}}, \bibinfo {author} {\bibfnamefont {C.}~\bibnamefont {Deprez}},
  \bibinfo {author} {\bibfnamefont {H.}~\bibnamefont {Toida}}, \bibinfo
  {author} {\bibfnamefont {K.}~\bibnamefont {Semba}}, \bibinfo {author}
  {\bibfnamefont {H.}~\bibnamefont {Yamaguchi}}, \bibinfo {author}
  {\bibfnamefont {W.~J.}\ \bibnamefont {Munro}}, \ and\ \bibinfo {author}
  {\bibfnamefont {S.}~\bibnamefont {Saito}},\ }\href@noop {} {\enquote
  {\bibinfo {title} {Observation of collective coupling between an engineered
  ensemble of macroscopic artificial atoms and a superconducting resonator},}\
  } (\bibinfo {year} {2016}),\ \bibinfo {note} {arXiv:1606.04222}\BibitemShut
  {NoStop}%
\bibitem [{\citenamefont {Buluta}\ and\ \citenamefont
  {Nori}(2009)}]{Buluta108}%
  \BibitemOpen
  \bibfield  {author} {\bibinfo {author} {\bibfnamefont {I.}~\bibnamefont
  {Buluta}}\ and\ \bibinfo {author} {\bibfnamefont {F.}~\bibnamefont {Nori}},\
  }\bibfield  {title} {\enquote {\bibinfo {title} {Quantum simulators},}\
  }\href {\doibase 10.1126/science.1177838} {\bibfield  {journal} {\bibinfo
  {journal} {Science}\ }\textbf {\bibinfo {volume} {326}},\ \bibinfo {pages}
  {108--111} (\bibinfo {year} {2009})}\BibitemShut {NoStop}%
\bibitem [{\citenamefont {Georgescu}\ \emph {et~al.}(2014)\citenamefont
  {Georgescu}, \citenamefont {Ashhab},\ and\ \citenamefont {Nori}}]{rmp2014}%
  \BibitemOpen
  \bibfield  {author} {\bibinfo {author} {\bibfnamefont {I.~M.}\ \bibnamefont
  {Georgescu}}, \bibinfo {author} {\bibfnamefont {S.}~\bibnamefont {Ashhab}}, \
  and\ \bibinfo {author} {\bibfnamefont {Franco}\ \bibnamefont {Nori}},\
  }\bibfield  {title} {\enquote {\bibinfo {title} {Quantum simulation},}\
  }\href {\doibase 10.1103/RevModPhys.86.153} {\bibfield  {journal} {\bibinfo
  {journal} {Rev. Mod. Phys.}\ }\textbf {\bibinfo {volume} {86}},\ \bibinfo
  {pages} {153--185} (\bibinfo {year} {2014})}\BibitemShut {NoStop}%
\bibitem [{\citenamefont {Hepp}\ and\ \citenamefont
  {Lieb}(1973)}]{hepp1973superradiant}%
  \BibitemOpen
  \bibfield  {author} {\bibinfo {author} {\bibfnamefont {K.}~\bibnamefont
  {Hepp}}\ and\ \bibinfo {author} {\bibfnamefont {E.~H.}\ \bibnamefont
  {Lieb}},\ }\bibfield  {title} {\enquote {\bibinfo {title} {On the
  superradiant phase transition for molecules in a quantized radiation field:
  The dicke maser model},}\ }\href@noop {} {\bibfield  {journal} {\bibinfo
  {journal} {Annals of Physics}\ }\textbf {\bibinfo {volume} {76}},\ \bibinfo
  {pages} {360--404} (\bibinfo {year} {1973})}\BibitemShut {NoStop}%
\bibitem [{\citenamefont {Wang}\ and\ \citenamefont
  {Hioe}(1973)}]{wang1973phase}%
  \BibitemOpen
  \bibfield  {author} {\bibinfo {author} {\bibfnamefont {Y.~K.}\ \bibnamefont
  {Wang}}\ and\ \bibinfo {author} {\bibfnamefont {F.~T.}\ \bibnamefont
  {Hioe}},\ }\bibfield  {title} {\enquote {\bibinfo {title} {Phase transition
  in the dicke model of superradiance},}\ }\href@noop {} {\bibfield  {journal}
  {\bibinfo  {journal} {Phys. Rev. A}\ }\textbf {\bibinfo {volume} {7}},\
  \bibinfo {pages} {831} (\bibinfo {year} {1973})}\BibitemShut {NoStop}%
\bibitem [{\citenamefont {Emary}\ and\ \citenamefont
  {Brandes}(2003)}]{emary2003chaos}%
  \BibitemOpen
  \bibfield  {author} {\bibinfo {author} {\bibfnamefont {C.}~\bibnamefont
  {Emary}}\ and\ \bibinfo {author} {\bibfnamefont {T.}~\bibnamefont
  {Brandes}},\ }\bibfield  {title} {\enquote {\bibinfo {title} {Chaos and the
  quantum phase transition in the dicke model},}\ }\href@noop {} {\bibfield
  {journal} {\bibinfo  {journal} {Phys. Rev. E}\ }\textbf {\bibinfo {volume}
  {67}},\ \bibinfo {pages} {066203} (\bibinfo {year} {2003})}\BibitemShut
  {NoStop}%
\bibitem [{\citenamefont {Lambert}\ \emph {et~al.}(2009)\citenamefont
  {Lambert}, \citenamefont {Chen}, \citenamefont {Johansson},\ and\
  \citenamefont {Nori}}]{lambert2009quantum}%
  \BibitemOpen
  \bibfield  {author} {\bibinfo {author} {\bibfnamefont {N.}~\bibnamefont
  {Lambert}}, \bibinfo {author} {\bibfnamefont {Y.~N.}\ \bibnamefont {Chen}},
  \bibinfo {author} {\bibfnamefont {R.}~\bibnamefont {Johansson}}, \ and\
  \bibinfo {author} {\bibfnamefont {F.}~\bibnamefont {Nori}},\ }\bibfield
  {title} {\enquote {\bibinfo {title} {Quantum chaos and critical behavior on a
  chip},}\ }\href@noop {} {\bibfield  {journal} {\bibinfo  {journal} {Phys.
  Rev. B}\ }\textbf {\bibinfo {volume} {80}},\ \bibinfo {pages} {165308}
  (\bibinfo {year} {2009})}\BibitemShut {NoStop}%
\bibitem [{\citenamefont {Lambert}\ \emph {et~al.}(2004)\citenamefont
  {Lambert}, \citenamefont {Emary},\ and\ \citenamefont
  {Brandes}}]{lambert2004entanglement}%
  \BibitemOpen
  \bibfield  {author} {\bibinfo {author} {\bibfnamefont {N.}~\bibnamefont
  {Lambert}}, \bibinfo {author} {\bibfnamefont {C.}~\bibnamefont {Emary}}, \
  and\ \bibinfo {author} {\bibfnamefont {T.}~\bibnamefont {Brandes}},\
  }\bibfield  {title} {\enquote {\bibinfo {title} {Entanglement and the phase
  transition in single-mode superradiance},}\ }\href@noop {} {\bibfield
  {journal} {\bibinfo  {journal} {Phys. Rev. Lett.}\ }\textbf {\bibinfo
  {volume} {92}},\ \bibinfo {pages} {073602} (\bibinfo {year}
  {2004})}\BibitemShut {NoStop}%
\bibitem [{\citenamefont {Rakhmanov}\ \emph {et~al.}(2008)\citenamefont
  {Rakhmanov}, \citenamefont {Zagoskin}, \citenamefont {Savel'ev},\ and\
  \citenamefont {Nori}}]{PhysRevB.77.144507}%
  \BibitemOpen
  \bibfield  {author} {\bibinfo {author} {\bibfnamefont {A.~L.}\ \bibnamefont
  {Rakhmanov}}, \bibinfo {author} {\bibfnamefont {A.~M.}\ \bibnamefont
  {Zagoskin}}, \bibinfo {author} {\bibfnamefont {S.}~\bibnamefont {Savel'ev}},
  \ and\ \bibinfo {author} {\bibfnamefont {F.}~\bibnamefont {Nori}},\
  }\bibfield  {title} {\enquote {\bibinfo {title} {Quantum metamaterials:
  {E}lectromagnetic waves in a {J}osephson qubit line},}\ }\href {\doibase
  10.1103/PhysRevB.77.144507} {\bibfield  {journal} {\bibinfo  {journal} {Phys.
  Rev. B}\ }\textbf {\bibinfo {volume} {77}},\ \bibinfo {pages} {144507}
  (\bibinfo {year} {2008})}\BibitemShut {NoStop}%
\bibitem [{\citenamefont {Soukoulis}\ and\ \citenamefont
  {Wegener}(2011)}]{soukoulis2011past}%
  \BibitemOpen
  \bibfield  {author} {\bibinfo {author} {\bibfnamefont {C.~M.}\ \bibnamefont
  {Soukoulis}}\ and\ \bibinfo {author} {\bibfnamefont {M.}~\bibnamefont
  {Wegener}},\ }\bibfield  {title} {\enquote {\bibinfo {title} {Past
  achievements and future challenges in the development of three-dimensional
  photonic metamaterials},}\ }\href@noop {} {\bibfield  {journal} {\bibinfo
  {journal} {Nature Photonics}\ }\textbf {\bibinfo {volume} {5}},\ \bibinfo
  {pages} {523--530} (\bibinfo {year} {2011})}\BibitemShut {NoStop}%
\bibitem [{\citenamefont {Zheludev}\ and\ \citenamefont
  {Kivshar}(2012)}]{zheludev2012metamaterials}%
  \BibitemOpen
  \bibfield  {author} {\bibinfo {author} {\bibfnamefont {N.~I.}\ \bibnamefont
  {Zheludev}}\ and\ \bibinfo {author} {\bibfnamefont {Y.~S.}\ \bibnamefont
  {Kivshar}},\ }\bibfield  {title} {\enquote {\bibinfo {title} {From
  metamaterials to metadevices},}\ }\href@noop {} {\bibfield  {journal}
  {\bibinfo  {journal} {Nature Materials}\ }\textbf {\bibinfo {volume} {11}},\
  \bibinfo {pages} {917--924} (\bibinfo {year} {2012})}\BibitemShut {NoStop}%
\bibitem [{\citenamefont {Macha}\ \emph {et~al.}(2014)\citenamefont {Macha},
  \citenamefont {Oelsner}, \citenamefont {Reiner}, \citenamefont {Marthaler},
  \citenamefont {Andr{\'e}}, \citenamefont {Sch{\"o}n}, \citenamefont
  {H{\"u}bner}, \citenamefont {Meyer}, \citenamefont {llIichev},\ and\
  \citenamefont {Ustinov}}]{macha2014implementation}%
  \BibitemOpen
  \bibfield  {author} {\bibinfo {author} {\bibfnamefont {P.}~\bibnamefont
  {Macha}}, \bibinfo {author} {\bibfnamefont {G.}~\bibnamefont {Oelsner}},
  \bibinfo {author} {\bibfnamefont {J.~M.}\ \bibnamefont {Reiner}}, \bibinfo
  {author} {\bibfnamefont {M.}~\bibnamefont {Marthaler}}, \bibinfo {author}
  {\bibfnamefont {S.}~\bibnamefont {Andr{\'e}}}, \bibinfo {author}
  {\bibfnamefont {G.}~\bibnamefont {Sch{\"o}n}}, \bibinfo {author}
  {\bibfnamefont {U.}~\bibnamefont {H{\"u}bner}}, \bibinfo {author}
  {\bibfnamefont {H.~G.}\ \bibnamefont {Meyer}}, \bibinfo {author}
  {\bibfnamefont {E.}~\bibnamefont {llIichev}}, \ and\ \bibinfo {author}
  {\bibfnamefont {A.~V.}\ \bibnamefont {Ustinov}},\ }\bibfield  {title}
  {\enquote {\bibinfo {title} {Implementation of a quantum metamaterial using
  superconducting qubits},}\ }\href@noop {} {\bibfield  {journal} {\bibinfo
  {journal} {Nature Comms}\ }\textbf {\bibinfo {volume} {5}} (\bibinfo {year}
  {2014})}\BibitemShut {NoStop}%
\bibitem [{\citenamefont {Kitagawa}\ and\ \citenamefont
  {Ueda}(1993)}]{kitagawa1993squeezed}%
  \BibitemOpen
  \bibfield  {author} {\bibinfo {author} {\bibfnamefont {M.}~\bibnamefont
  {Kitagawa}}\ and\ \bibinfo {author} {\bibfnamefont {M.}~\bibnamefont
  {Ueda}},\ }\bibfield  {title} {\enquote {\bibinfo {title} {Squeezed spin
  states},}\ }\href@noop {} {\bibfield  {journal} {\bibinfo  {journal} {Phys.
  Rev. A}\ }\textbf {\bibinfo {volume} {47}},\ \bibinfo {pages} {5138}
  (\bibinfo {year} {1993})}\BibitemShut {NoStop}%
\bibitem [{\citenamefont {Ma}\ \emph {et~al.}(2011)\citenamefont {Ma},
  \citenamefont {Wang}, \citenamefont {Sun},\ and\ \citenamefont
  {Nori}}]{Ma2011}%
  \BibitemOpen
  \bibfield  {author} {\bibinfo {author} {\bibfnamefont {J.}~\bibnamefont
  {Ma}}, \bibinfo {author} {\bibfnamefont {X.}~\bibnamefont {Wang}}, \bibinfo
  {author} {\bibfnamefont {C.P.}\ \bibnamefont {Sun}}, \ and\ \bibinfo {author}
  {\bibfnamefont {F.}~\bibnamefont {Nori}},\ }\bibfield  {title} {\enquote
  {\bibinfo {title} {Quantum spin squeezing},}\ }\href {\doibase
  http://dx.doi.org/10.1016/j.physrep.2011.08.003} {\bibfield  {journal}
  {\bibinfo  {journal} {Physics Reports}\ }\textbf {\bibinfo {volume} {509}},\
  \bibinfo {pages} {89 -- 165} (\bibinfo {year} {2011})}\BibitemShut {NoStop}%
\bibitem [{\citenamefont {Bennett}\ \emph {et~al.}(2013)\citenamefont
  {Bennett}, \citenamefont {Yao}, \citenamefont {Otterbach}, \citenamefont
  {Zoller}, \citenamefont {Rabl},\ and\ \citenamefont
  {Lukin}}]{bennett2013phonon}%
  \BibitemOpen
  \bibfield  {author} {\bibinfo {author} {\bibfnamefont {S.~D.}\ \bibnamefont
  {Bennett}}, \bibinfo {author} {\bibfnamefont {N.~Y.}\ \bibnamefont {Yao}},
  \bibinfo {author} {\bibfnamefont {J.}~\bibnamefont {Otterbach}}, \bibinfo
  {author} {\bibfnamefont {P.}~\bibnamefont {Zoller}}, \bibinfo {author}
  {\bibfnamefont {P.}~\bibnamefont {Rabl}}, \ and\ \bibinfo {author}
  {\bibfnamefont {M.~D.}\ \bibnamefont {Lukin}},\ }\bibfield  {title} {\enquote
  {\bibinfo {title} {Phonon-induced spin-spin interactions in diamond
  nanostructures: application to spin squeezing},}\ }\href@noop {} {\bibfield
  {journal} {\bibinfo  {journal} {Phys. Rev. Lett.}\ }\textbf {\bibinfo
  {volume} {110}},\ \bibinfo {pages} {156402} (\bibinfo {year}
  {2013})}\BibitemShut {NoStop}%
\bibitem [{\citenamefont {Tanaka}\ \emph {et~al.}(2015)\citenamefont {Tanaka},
  \citenamefont {Knott}, \citenamefont {Matsuzaki}, \citenamefont {Dooley},
  \citenamefont {Yamaguchi}, \citenamefont {Munro},\ and\ \citenamefont
  {Saito}}]{tanaka2015proposed}%
  \BibitemOpen
  \bibfield  {author} {\bibinfo {author} {\bibfnamefont {T.}~\bibnamefont
  {Tanaka}}, \bibinfo {author} {\bibfnamefont {P.}~\bibnamefont {Knott}},
  \bibinfo {author} {\bibfnamefont {Y.}~\bibnamefont {Matsuzaki}}, \bibinfo
  {author} {\bibfnamefont {S.}~\bibnamefont {Dooley}}, \bibinfo {author}
  {\bibfnamefont {H.}~\bibnamefont {Yamaguchi}}, \bibinfo {author}
  {\bibfnamefont {W.~J.}\ \bibnamefont {Munro}}, \ and\ \bibinfo {author}
  {\bibfnamefont {S.}~\bibnamefont {Saito}},\ }\bibfield  {title} {\enquote
  {\bibinfo {title} {Proposed robust entanglement-based magnetic field sensor
  beyond the standard quantum limit},}\ }\href@noop {} {\bibfield  {journal}
  {\bibinfo  {journal} {Phys. Rev. Lett.}\ }\textbf {\bibinfo {volume} {115}},\
  \bibinfo {pages} {170801} (\bibinfo {year} {2015})}\BibitemShut {NoStop}%
\bibitem [{\citenamefont {Imamo{\u{g}}lu}(2009)}]{imamouglu2009cavity}%
  \BibitemOpen
  \bibfield  {author} {\bibinfo {author} {\bibfnamefont {A.}~\bibnamefont
  {Imamo{\u{g}}lu}},\ }\bibfield  {title} {\enquote {\bibinfo {title} {Cavity
  {QED} based on collective magnetic dipole coupling: spin ensembles as hybrid
  two-level systems},}\ }\href@noop {} {\bibfield  {journal} {\bibinfo
  {journal} {Phys. Rev. Lett.}\ }\textbf {\bibinfo {volume} {102}},\ \bibinfo
  {pages} {083602} (\bibinfo {year} {2009})}\BibitemShut {NoStop}%
\bibitem [{\citenamefont {Wesenberg}\ \emph {et~al.}(2009)\citenamefont
  {Wesenberg}, \citenamefont {Ardavan}, \citenamefont {Briggs}, \citenamefont
  {Morton}, \citenamefont {Schoelkopf}, \citenamefont {Schuster},\ and\
  \citenamefont {M{\o}lmer}}]{wesenberg2009quantum}%
  \BibitemOpen
  \bibfield  {author} {\bibinfo {author} {\bibfnamefont {J.~H.}\ \bibnamefont
  {Wesenberg}}, \bibinfo {author} {\bibfnamefont {A.}~\bibnamefont {Ardavan}},
  \bibinfo {author} {\bibfnamefont {G.~A.~D}\ \bibnamefont {Briggs}}, \bibinfo
  {author} {\bibfnamefont {J.~J.~L.}\ \bibnamefont {Morton}}, \bibinfo {author}
  {\bibfnamefont {R.~J.}\ \bibnamefont {Schoelkopf}}, \bibinfo {author}
  {\bibfnamefont {D.~I.}\ \bibnamefont {Schuster}}, \ and\ \bibinfo {author}
  {\bibfnamefont {K.}~\bibnamefont {M{\o}lmer}},\ }\bibfield  {title} {\enquote
  {\bibinfo {title} {{Quantum computing with an electron spin ensemble}},}\
  }\href@noop {} {\bibfield  {journal} {\bibinfo  {journal} {Phys. Rev. Lett.}\
  }\textbf {\bibinfo {volume} {103}},\ \bibinfo {pages} {70502} (\bibinfo
  {year} {2009})}\BibitemShut {NoStop}%
\bibitem [{\citenamefont {Xiang}\ \emph {et~al.}(2013)\citenamefont {Xiang},
  \citenamefont {Ashhab}, \citenamefont {You},\ and\ \citenamefont
  {Nori}}]{rmp2013hybrids}%
  \BibitemOpen
  \bibfield  {author} {\bibinfo {author} {\bibfnamefont {Z-L.}\ \bibnamefont
  {Xiang}}, \bibinfo {author} {\bibfnamefont {S.}~\bibnamefont {Ashhab}},
  \bibinfo {author} {\bibfnamefont {J.~Q.}\ \bibnamefont {You}}, \ and\
  \bibinfo {author} {\bibfnamefont {F.}~\bibnamefont {Nori}},\ }\bibfield
  {title} {\enquote {\bibinfo {title} {Hybrid quantum circuits: Superconducting
  circuits interacting with other quantum systems},}\ }\href {\doibase
  10.1103/RevModPhys.85.623} {\bibfield  {journal} {\bibinfo  {journal} {Rev.
  Mod. Phys.}\ }\textbf {\bibinfo {volume} {85}},\ \bibinfo {pages} {623--653}
  (\bibinfo {year} {2013})}\BibitemShut {NoStop}%
\bibitem [{\citenamefont {Schuster}\ \emph {et~al.}(2010)\citenamefont
  {Schuster}, \citenamefont {Sears}, \citenamefont {Ginossar}, \citenamefont
  {DiCarlo}, \citenamefont {Frunzio}, \citenamefont {Morton}, \citenamefont
  {Wu}, \citenamefont {Briggs}, \citenamefont {Buckley}, \citenamefont
  {Awschalom},\ and\ \citenamefont {Schoelkopf}}]{schuster2010high}%
  \BibitemOpen
  \bibfield  {author} {\bibinfo {author} {\bibfnamefont {D.~I.}\ \bibnamefont
  {Schuster}}, \bibinfo {author} {\bibfnamefont {A.~P.}\ \bibnamefont {Sears}},
  \bibinfo {author} {\bibfnamefont {E.}~\bibnamefont {Ginossar}}, \bibinfo
  {author} {\bibfnamefont {L.}~\bibnamefont {DiCarlo}}, \bibinfo {author}
  {\bibfnamefont {L.}~\bibnamefont {Frunzio}}, \bibinfo {author} {\bibfnamefont
  {J.~J.~L.}\ \bibnamefont {Morton}}, \bibinfo {author} {\bibfnamefont
  {H.}~\bibnamefont {Wu}}, \bibinfo {author} {\bibfnamefont {G.~A.~D.}\
  \bibnamefont {Briggs}}, \bibinfo {author} {\bibfnamefont {B.~B.}\
  \bibnamefont {Buckley}}, \bibinfo {author} {\bibfnamefont {D.~D.}\
  \bibnamefont {Awschalom}}, \ and\ \bibinfo {author} {\bibfnamefont {R.~J.}\
  \bibnamefont {Schoelkopf}},\ }\bibfield  {title} {\enquote {\bibinfo {title}
  {{High-Cooperativity Coupling of Electron-Spin Ensembles to Superconducting
  Cavities}},}\ }\href@noop {} {\bibfield  {journal} {\bibinfo  {journal}
  {Phys. Rev. Lett.}\ }\textbf {\bibinfo {volume} {105}},\ \bibinfo {pages}
  {140501} (\bibinfo {year} {2010})}\BibitemShut {NoStop}%
\bibitem [{\citenamefont {Wu}\ \emph {et~al.}(2010)\citenamefont {Wu},
  \citenamefont {George}, \citenamefont {Wesenberg}, \citenamefont {M{\o}lmer},
  \citenamefont {Schuster}, \citenamefont {Schoelkopf}, \citenamefont {Itoh},
  \citenamefont {Ardavan}, \citenamefont {Morton},\ and\ \citenamefont
  {Briggs}}]{wu2010storage}%
  \BibitemOpen
  \bibfield  {author} {\bibinfo {author} {\bibfnamefont {H.}~\bibnamefont
  {Wu}}, \bibinfo {author} {\bibfnamefont {R.~E.}\ \bibnamefont {George}},
  \bibinfo {author} {\bibfnamefont {J.~H.}\ \bibnamefont {Wesenberg}}, \bibinfo
  {author} {\bibfnamefont {K.}~\bibnamefont {M{\o}lmer}}, \bibinfo {author}
  {\bibfnamefont {D.~I.}\ \bibnamefont {Schuster}}, \bibinfo {author}
  {\bibfnamefont {R.~J.}\ \bibnamefont {Schoelkopf}}, \bibinfo {author}
  {\bibfnamefont {K.~M.}\ \bibnamefont {Itoh}}, \bibinfo {author}
  {\bibfnamefont {A.}~\bibnamefont {Ardavan}}, \bibinfo {author} {\bibfnamefont
  {J.~J.~L.}\ \bibnamefont {Morton}}, \ and\ \bibinfo {author} {\bibfnamefont
  {G.~A.~D.}\ \bibnamefont {Briggs}},\ }\bibfield  {title} {\enquote {\bibinfo
  {title} {Storage of multiple coherent microwave excitations in an electron
  spin ensemble},}\ }\href@noop {} {\bibfield  {journal} {\bibinfo  {journal}
  {Phys. Rev. Lett.}\ }\textbf {\bibinfo {volume} {105}},\ \bibinfo {pages}
  {140503} (\bibinfo {year} {2010})}\BibitemShut {NoStop}%
\bibitem [{\citenamefont {Kubo}\ \emph {et~al.}(2010)\citenamefont {Kubo},
  \citenamefont {Ong}, \citenamefont {Bertet}, \citenamefont {Vion},
  \citenamefont {Jacques}, \citenamefont {Zheng}, \citenamefont {Dr{\'e}au},
  \citenamefont {Roch}, \citenamefont {Auffeves}, \citenamefont {Jelezko},
  \citenamefont {Wrachtrup}, \citenamefont {Barthe}, \citenamefont {Bergonzo},\
  and\ \citenamefont {Esteve}}]{kubo2010strong}%
  \BibitemOpen
  \bibfield  {author} {\bibinfo {author} {\bibfnamefont {Y.}~\bibnamefont
  {Kubo}}, \bibinfo {author} {\bibfnamefont {F.~R.}\ \bibnamefont {Ong}},
  \bibinfo {author} {\bibfnamefont {P.}~\bibnamefont {Bertet}}, \bibinfo
  {author} {\bibfnamefont {D.}~\bibnamefont {Vion}}, \bibinfo {author}
  {\bibfnamefont {V.}~\bibnamefont {Jacques}}, \bibinfo {author} {\bibfnamefont
  {D.}~\bibnamefont {Zheng}}, \bibinfo {author} {\bibfnamefont
  {A.}~\bibnamefont {Dr{\'e}au}}, \bibinfo {author} {\bibfnamefont {J.~F.}\
  \bibnamefont {Roch}}, \bibinfo {author} {\bibfnamefont {A.}~\bibnamefont
  {Auffeves}}, \bibinfo {author} {\bibfnamefont {F.}~\bibnamefont {Jelezko}},
  \bibinfo {author} {\bibfnamefont {J.}~\bibnamefont {Wrachtrup}}, \bibinfo
  {author} {\bibfnamefont {M.~F.}\ \bibnamefont {Barthe}}, \bibinfo {author}
  {\bibfnamefont {P.}~\bibnamefont {Bergonzo}}, \ and\ \bibinfo {author}
  {\bibfnamefont {D.}~\bibnamefont {Esteve}},\ }\bibfield  {title} {\enquote
  {\bibinfo {title} {{Strong Coupling of a Spin Ensemble to a Superconducting
  Resonator}},}\ }\href@noop {} {\bibfield  {journal} {\bibinfo  {journal}
  {Phys. Rev. Lett.}\ }\textbf {\bibinfo {volume} {105}},\ \bibinfo {pages}
  {140502} (\bibinfo {year} {2010})}\BibitemShut {NoStop}%
\bibitem [{\citenamefont {Ams{\"u}ss}\ \emph {et~al.}(2011)\citenamefont
  {Ams{\"u}ss}, \citenamefont {Koller}, \citenamefont {N{\"o}bauer},
  \citenamefont {Putz}, \citenamefont {Rotter}, \citenamefont {Sandner},
  \citenamefont {Schneider}, \citenamefont {Schramb{\"o}ck}, \citenamefont
  {Steinhauser}, \citenamefont {Ritsch}, \citenamefont {Schmiedmayer},\ and\
  \citenamefont {Majer}}]{amsuss2011cavity}%
  \BibitemOpen
  \bibfield  {author} {\bibinfo {author} {\bibfnamefont {R.}~\bibnamefont
  {Ams{\"u}ss}}, \bibinfo {author} {\bibfnamefont {C.~H.}\ \bibnamefont
  {Koller}}, \bibinfo {author} {\bibfnamefont {T.}~\bibnamefont {N{\"o}bauer}},
  \bibinfo {author} {\bibfnamefont {S.}~\bibnamefont {Putz}}, \bibinfo {author}
  {\bibfnamefont {S.}~\bibnamefont {Rotter}}, \bibinfo {author} {\bibfnamefont
  {K.}~\bibnamefont {Sandner}}, \bibinfo {author} {\bibfnamefont
  {S.}~\bibnamefont {Schneider}}, \bibinfo {author} {\bibfnamefont
  {M.}~\bibnamefont {Schramb{\"o}ck}}, \bibinfo {author} {\bibfnamefont
  {G.}~\bibnamefont {Steinhauser}}, \bibinfo {author} {\bibfnamefont
  {H.}~\bibnamefont {Ritsch}}, \bibinfo {author} {\bibfnamefont
  {J.}~\bibnamefont {Schmiedmayer}}, \ and\ \bibinfo {author} {\bibfnamefont
  {J.}~\bibnamefont {Majer}},\ }\bibfield  {title} {\enquote {\bibinfo {title}
  {Cavity {QED} with magnetically coupled collective spin states},}\
  }\href@noop {} {\bibfield  {journal} {\bibinfo  {journal} {Phys. Rev. Lett.}\
  }\textbf {\bibinfo {volume} {107}},\ \bibinfo {pages} {060502} (\bibinfo
  {year} {2011})}\BibitemShut {NoStop}%
\bibitem [{\citenamefont {Kubo}\ \emph {et~al.}(2011)\citenamefont {Kubo},
  \citenamefont {Grezes}, \citenamefont {Dewes}, \citenamefont {Umeda},
  \citenamefont {Isoya}, \citenamefont {Sumiya}, \citenamefont {Morishita},
  \citenamefont {Abe}, \citenamefont {Onoda}, \citenamefont {Ohshima},
  \citenamefont {Jacques}, \citenamefont {Dreau}, \citenamefont {Roch},
  \citenamefont {Diniz}, \citenamefont {Auffeves}, \citenamefont {Vion},
  \citenamefont {Esteve},\ and\ \citenamefont {Bertet}}]{kubo2011hybrid}%
  \BibitemOpen
  \bibfield  {author} {\bibinfo {author} {\bibfnamefont {Y.}~\bibnamefont
  {Kubo}}, \bibinfo {author} {\bibfnamefont {C.}~\bibnamefont {Grezes}},
  \bibinfo {author} {\bibfnamefont {A.}~\bibnamefont {Dewes}}, \bibinfo
  {author} {\bibfnamefont {T.}~\bibnamefont {Umeda}}, \bibinfo {author}
  {\bibfnamefont {J.}~\bibnamefont {Isoya}}, \bibinfo {author} {\bibfnamefont
  {H.}~\bibnamefont {Sumiya}}, \bibinfo {author} {\bibfnamefont
  {N.}~\bibnamefont {Morishita}}, \bibinfo {author} {\bibfnamefont
  {H.}~\bibnamefont {Abe}}, \bibinfo {author} {\bibfnamefont {S.}~\bibnamefont
  {Onoda}}, \bibinfo {author} {\bibfnamefont {T.}~\bibnamefont {Ohshima}},
  \bibinfo {author} {\bibfnamefont {V.}~\bibnamefont {Jacques}}, \bibinfo
  {author} {\bibfnamefont {A.}~\bibnamefont {Dreau}}, \bibinfo {author}
  {\bibfnamefont {J.~F.}\ \bibnamefont {Roch}}, \bibinfo {author}
  {\bibfnamefont {I.}~\bibnamefont {Diniz}}, \bibinfo {author} {\bibfnamefont
  {A.}~\bibnamefont {Auffeves}}, \bibinfo {author} {\bibfnamefont
  {D.}~\bibnamefont {Vion}}, \bibinfo {author} {\bibfnamefont {D.}~\bibnamefont
  {Esteve}}, \ and\ \bibinfo {author} {\bibfnamefont {P.}~\bibnamefont
  {Bertet}},\ }\bibfield  {title} {\enquote {\bibinfo {title} {Hybrid quantum
  circuit with a superconducting qubit coupled to a spin ensemble},}\
  }\href@noop {} {\bibfield  {journal} {\bibinfo  {journal} {Phys. Rev. Lett.}\
  }\textbf {\bibinfo {volume} {107}},\ \bibinfo {pages} {220501} (\bibinfo
  {year} {2011})}\BibitemShut {NoStop}%
\bibitem [{\citenamefont {Zhu}\ \emph {et~al.}(2011)\citenamefont {Zhu},
  \citenamefont {Saito}, \citenamefont {Kemp}, \citenamefont {Kakuyanagi},
  \citenamefont {Karimoto}, \citenamefont {Nakano}, \citenamefont {Munro},
  \citenamefont {Tokura}, \citenamefont {Everitt}, \citenamefont {Nemoto},
  \citenamefont {K.}, \citenamefont {M.},\ and\ \citenamefont
  {Semba}}]{zhu2011coherent}%
  \BibitemOpen
  \bibfield  {author} {\bibinfo {author} {\bibfnamefont {X.}~\bibnamefont
  {Zhu}}, \bibinfo {author} {\bibfnamefont {S.}~\bibnamefont {Saito}}, \bibinfo
  {author} {\bibfnamefont {A.}~\bibnamefont {Kemp}}, \bibinfo {author}
  {\bibfnamefont {K.}~\bibnamefont {Kakuyanagi}}, \bibinfo {author}
  {\bibfnamefont {S.}~\bibnamefont {Karimoto}}, \bibinfo {author}
  {\bibfnamefont {H.}~\bibnamefont {Nakano}}, \bibinfo {author} {\bibfnamefont
  {W.~J.}\ \bibnamefont {Munro}}, \bibinfo {author} {\bibfnamefont
  {Y.}~\bibnamefont {Tokura}}, \bibinfo {author} {\bibfnamefont {M.~S.}\
  \bibnamefont {Everitt}}, \bibinfo {author} {\bibfnamefont {K.}~\bibnamefont
  {Nemoto}}, \bibinfo {author} {\bibfnamefont {Makoto}\ \bibnamefont {K.}},
  \bibinfo {author} {\bibfnamefont {Norikazu}\ \bibnamefont {M.}}, \ and\
  \bibinfo {author} {\bibfnamefont {K.}~\bibnamefont {Semba}},\ }\bibfield
  {title} {\enquote {\bibinfo {title} {Coherent coupling of a superconducting
  flux qubit to an electron spin ensemble in diamond},}\ }\href@noop {}
  {\bibfield  {journal} {\bibinfo  {journal} {Nature}\ }\textbf {\bibinfo
  {volume} {478}},\ \bibinfo {pages} {221--224} (\bibinfo {year}
  {2011})}\BibitemShut {NoStop}%
\bibitem [{\citenamefont {Kubo}\ \emph
  {et~al.}(2012{\natexlab{a}})\citenamefont {Kubo}, \citenamefont {Diniz},
  \citenamefont {Dewes}, \citenamefont {Jacques}, \citenamefont {Dr{\'e}au},
  \citenamefont {Roch}, \citenamefont {Auff{\`e}ves}, \citenamefont {Vion},
  \citenamefont {Esteve},\ and\ \citenamefont {Bertet}}]{kubo2012storage}%
  \BibitemOpen
  \bibfield  {author} {\bibinfo {author} {\bibfnamefont {Y.}~\bibnamefont
  {Kubo}}, \bibinfo {author} {\bibfnamefont {I.}~\bibnamefont {Diniz}},
  \bibinfo {author} {\bibfnamefont {A.}~\bibnamefont {Dewes}}, \bibinfo
  {author} {\bibfnamefont {V.}~\bibnamefont {Jacques}}, \bibinfo {author}
  {\bibfnamefont {A.}~\bibnamefont {Dr{\'e}au}}, \bibinfo {author}
  {\bibfnamefont {J.~F.}\ \bibnamefont {Roch}}, \bibinfo {author}
  {\bibfnamefont {A.}~\bibnamefont {Auff{\`e}ves}}, \bibinfo {author}
  {\bibfnamefont {D.}~\bibnamefont {Vion}}, \bibinfo {author} {\bibfnamefont
  {D.}~\bibnamefont {Esteve}}, \ and\ \bibinfo {author} {\bibfnamefont
  {P.}~\bibnamefont {Bertet}},\ }\bibfield  {title} {\enquote {\bibinfo {title}
  {Storage and retrieval of a microwave field in a spin ensemble},}\
  }\href@noop {} {\bibfield  {journal} {\bibinfo  {journal} {Phys. Rev. A.}\
  }\textbf {\bibinfo {volume} {85}},\ \bibinfo {pages} {012333} (\bibinfo
  {year} {2012}{\natexlab{a}})}\BibitemShut {NoStop}%
\bibitem [{\citenamefont {Kubo}\ \emph
  {et~al.}(2012{\natexlab{b}})\citenamefont {Kubo}, \citenamefont {Diniz},
  \citenamefont {Grezes}, \citenamefont {Umeda}, \citenamefont {Isoya},
  \citenamefont {Sumiya}, \citenamefont {Yamamoto}, \citenamefont {Abe},
  \citenamefont {Onoda}, \citenamefont {Ohshima}, \citenamefont {Jacques},
  \citenamefont {Dr{\'e}au}, \citenamefont {Roch}, \citenamefont
  {Auff{\`e}ves}, \citenamefont {Vion}, \citenamefont {Esteve},\ and\
  \citenamefont {Bertet}}]{kubo2012electron}%
  \BibitemOpen
  \bibfield  {author} {\bibinfo {author} {\bibfnamefont {Y.}~\bibnamefont
  {Kubo}}, \bibinfo {author} {\bibfnamefont {I.}~\bibnamefont {Diniz}},
  \bibinfo {author} {\bibfnamefont {C.}~\bibnamefont {Grezes}}, \bibinfo
  {author} {\bibfnamefont {T.}~\bibnamefont {Umeda}}, \bibinfo {author}
  {\bibfnamefont {J.}~\bibnamefont {Isoya}}, \bibinfo {author} {\bibfnamefont
  {H.}~\bibnamefont {Sumiya}}, \bibinfo {author} {\bibfnamefont
  {T.}~\bibnamefont {Yamamoto}}, \bibinfo {author} {\bibfnamefont
  {H.}~\bibnamefont {Abe}}, \bibinfo {author} {\bibfnamefont {S.}~\bibnamefont
  {Onoda}}, \bibinfo {author} {\bibfnamefont {T.}~\bibnamefont {Ohshima}},
  \bibinfo {author} {\bibfnamefont {V.}~\bibnamefont {Jacques}}, \bibinfo
  {author} {\bibfnamefont {A.}~\bibnamefont {Dr{\'e}au}}, \bibinfo {author}
  {\bibfnamefont {J.~F.}\ \bibnamefont {Roch}}, \bibinfo {author}
  {\bibfnamefont {A.}~\bibnamefont {Auff{\`e}ves}}, \bibinfo {author}
  {\bibfnamefont {D.}~\bibnamefont {Vion}}, \bibinfo {author} {\bibfnamefont
  {D.}~\bibnamefont {Esteve}}, \ and\ \bibinfo {author} {\bibfnamefont
  {P.}~\bibnamefont {Bertet}},\ }\bibfield  {title} {\enquote {\bibinfo {title}
  {Electron spin resonance detected by a superconducting qubit},}\ }\href@noop
  {} {\bibfield  {journal} {\bibinfo  {journal} {Phys. Rev. B}\ }\textbf
  {\bibinfo {volume} {86}},\ \bibinfo {pages} {064514} (\bibinfo {year}
  {2012}{\natexlab{b}})}\BibitemShut {NoStop}%
\bibitem [{\citenamefont {Marcos}\ \emph {et~al.}(2010)\citenamefont {Marcos},
  \citenamefont {Wubs}, \citenamefont {Taylor}, \citenamefont {Aguado},
  \citenamefont {Lukin},\ and\ \citenamefont
  {S{\o}rensen}}]{marcos2010coupling}%
  \BibitemOpen
  \bibfield  {author} {\bibinfo {author} {\bibfnamefont {D.}~\bibnamefont
  {Marcos}}, \bibinfo {author} {\bibfnamefont {M.}~\bibnamefont {Wubs}},
  \bibinfo {author} {\bibfnamefont {J.~M.}\ \bibnamefont {Taylor}}, \bibinfo
  {author} {\bibfnamefont {R.}~\bibnamefont {Aguado}}, \bibinfo {author}
  {\bibfnamefont {M.~D.}\ \bibnamefont {Lukin}}, \ and\ \bibinfo {author}
  {\bibfnamefont {A.~S.}\ \bibnamefont {S{\o}rensen}},\ }\bibfield  {title}
  {\enquote {\bibinfo {title} {{Coupling nitrogen-vacancy centers in diamond to
  superconducting flux qubits}},}\ }\href@noop {} {\bibfield  {journal}
  {\bibinfo  {journal} {Phys. Rev. Lett.}\ }\textbf {\bibinfo {volume} {105}},\
  \bibinfo {pages} {210501} (\bibinfo {year} {2010})}\BibitemShut {NoStop}%
\bibitem [{\citenamefont {Twamley}\ and\ \citenamefont
  {Barrett}(2010)}]{twamley2010superconducting}%
  \BibitemOpen
  \bibfield  {author} {\bibinfo {author} {\bibfnamefont {J.}~\bibnamefont
  {Twamley}}\ and\ \bibinfo {author} {\bibfnamefont {S.~D.}\ \bibnamefont
  {Barrett}},\ }\bibfield  {title} {\enquote {\bibinfo {title} {Superconducting
  cavity bus for single nitrogen-vacancy defect centers in diamond},}\
  }\href@noop {} {\bibfield  {journal} {\bibinfo  {journal} {Phys. Rev. B}\
  }\textbf {\bibinfo {volume} {81}},\ \bibinfo {pages} {241202} (\bibinfo
  {year} {2010})}\BibitemShut {NoStop}%
\bibitem [{\citenamefont {Matsuzaki}\ and\ \citenamefont
  {Nakano}(2012)}]{matsuzaki2012enhanced}%
  \BibitemOpen
  \bibfield  {author} {\bibinfo {author} {\bibfnamefont {Y.}~\bibnamefont
  {Matsuzaki}}\ and\ \bibinfo {author} {\bibfnamefont {H.}~\bibnamefont
  {Nakano}},\ }\bibfield  {title} {\enquote {\bibinfo {title} {Enhanced energy
  relaxation process of a quantum memory coupled to a superconducting qubit},}\
  }\href@noop {} {\bibfield  {journal} {\bibinfo  {journal} {Phys. Rev. B}\
  }\textbf {\bibinfo {volume} {86}},\ \bibinfo {pages} {184501} (\bibinfo
  {year} {2012})}\BibitemShut {NoStop}%
\bibitem [{\citenamefont {Saito}\ \emph {et~al.}(2013)\citenamefont {Saito},
  \citenamefont {Zhu}, \citenamefont {Ams{\"u}ss}, \citenamefont {Matsuzaki},
  \citenamefont {Kakuyanagi}, \citenamefont {Shimo-Oka}, \citenamefont
  {Mizuochi}, \citenamefont {Nemoto}, \citenamefont {Munro},\ and\
  \citenamefont {Semba}}]{saito2013towards}%
  \BibitemOpen
  \bibfield  {author} {\bibinfo {author} {\bibfnamefont {S.}~\bibnamefont
  {Saito}}, \bibinfo {author} {\bibfnamefont {X.}~\bibnamefont {Zhu}}, \bibinfo
  {author} {\bibfnamefont {R.}~\bibnamefont {Ams{\"u}ss}}, \bibinfo {author}
  {\bibfnamefont {Y.}~\bibnamefont {Matsuzaki}}, \bibinfo {author}
  {\bibfnamefont {K.}~\bibnamefont {Kakuyanagi}}, \bibinfo {author}
  {\bibfnamefont {T.}~\bibnamefont {Shimo-Oka}}, \bibinfo {author}
  {\bibfnamefont {N.}~\bibnamefont {Mizuochi}}, \bibinfo {author}
  {\bibfnamefont {K.}~\bibnamefont {Nemoto}}, \bibinfo {author} {\bibfnamefont
  {W.~J.}\ \bibnamefont {Munro}}, \ and\ \bibinfo {author} {\bibfnamefont
  {K.}~\bibnamefont {Semba}},\ }\bibfield  {title} {\enquote {\bibinfo {title}
  {Towards realizing a quantum memory for a superconducting qubit: Storage and
  retrieval of quantum states},}\ }\href@noop {} {\bibfield  {journal}
  {\bibinfo  {journal} {Phys. Rev. Lett.}\ }\textbf {\bibinfo {volume} {111}},\
  \bibinfo {pages} {107008} (\bibinfo {year} {2013})}\BibitemShut {NoStop}%
\bibitem [{\citenamefont {Julsgaard}\ \emph {et~al.}(2013)\citenamefont
  {Julsgaard}, \citenamefont {Grezes}, \citenamefont {Bertet},\ and\
  \citenamefont {M{\o}lmer}}]{julsgaard2013quantum}%
  \BibitemOpen
  \bibfield  {author} {\bibinfo {author} {\bibfnamefont {B.}~\bibnamefont
  {Julsgaard}}, \bibinfo {author} {\bibfnamefont {C.}~\bibnamefont {Grezes}},
  \bibinfo {author} {\bibfnamefont {P.}~\bibnamefont {Bertet}}, \ and\ \bibinfo
  {author} {\bibfnamefont {K.}~\bibnamefont {M{\o}lmer}},\ }\bibfield  {title}
  {\enquote {\bibinfo {title} {Quantum memory for microwave photons in an
  inhomogeneously broadened spin ensemble},}\ }\href@noop {} {\bibfield
  {journal} {\bibinfo  {journal} {Phys. Rev. Lett.}\ }\textbf {\bibinfo
  {volume} {110}},\ \bibinfo {pages} {250503} (\bibinfo {year}
  {2013})}\BibitemShut {NoStop}%
\bibitem [{\citenamefont {Diniz}\ \emph {et~al.}(2011)\citenamefont {Diniz},
  \citenamefont {Portolan}, \citenamefont {Ferreira}, \citenamefont
  {G{\'e}rard}, \citenamefont {Bertet},\ and\ \citenamefont
  {Auffeves}}]{diniz2011strongly}%
  \BibitemOpen
  \bibfield  {author} {\bibinfo {author} {\bibfnamefont {I.}~\bibnamefont
  {Diniz}}, \bibinfo {author} {\bibfnamefont {S.}~\bibnamefont {Portolan}},
  \bibinfo {author} {\bibfnamefont {R.}~\bibnamefont {Ferreira}}, \bibinfo
  {author} {\bibfnamefont {J.~M.}\ \bibnamefont {G{\'e}rard}}, \bibinfo
  {author} {\bibfnamefont {P.}~\bibnamefont {Bertet}}, \ and\ \bibinfo {author}
  {\bibfnamefont {A.}~\bibnamefont {Auffeves}},\ }\bibfield  {title} {\enquote
  {\bibinfo {title} {Strongly coupling a cavity to inhomogeneous ensembles of
  emitters: Potential for long-lived solid-state quantum memories},}\
  }\href@noop {} {\bibfield  {journal} {\bibinfo  {journal} {Phys. Rev. A}\
  }\textbf {\bibinfo {volume} {84}},\ \bibinfo {pages} {063810} (\bibinfo
  {year} {2011})}\BibitemShut {NoStop}%
\bibitem [{\citenamefont {Zhu}\ \emph {et~al.}(2014)\citenamefont {Zhu},
  \citenamefont {Matsuzaki}, \citenamefont {Amsuss}, \citenamefont
  {Kakuyanagi}, \citenamefont {Shimo-Oka}, \citenamefont {Mizuochi},
  \citenamefont {Nemoto}, \citenamefont {Munro}, \citenamefont {Semba},\ and\
  \citenamefont {Saito}}]{zhudark2014}%
  \BibitemOpen
  \bibfield  {author} {\bibinfo {author} {\bibfnamefont {X.}~\bibnamefont
  {Zhu}}, \bibinfo {author} {\bibfnamefont {Y.}~\bibnamefont {Matsuzaki}},
  \bibinfo {author} {\bibfnamefont {R.}~\bibnamefont {Amsuss}}, \bibinfo
  {author} {\bibfnamefont {K.}~\bibnamefont {Kakuyanagi}}, \bibinfo {author}
  {\bibfnamefont {T.}~\bibnamefont {Shimo-Oka}}, \bibinfo {author}
  {\bibfnamefont {N.}~\bibnamefont {Mizuochi}}, \bibinfo {author}
  {\bibfnamefont {K.}~\bibnamefont {Nemoto}}, \bibinfo {author} {\bibfnamefont
  {W.~J.}\ \bibnamefont {Munro}}, \bibinfo {author} {\bibfnamefont
  {K.}~\bibnamefont {Semba}}, \ and\ \bibinfo {author} {\bibfnamefont
  {S.}~\bibnamefont {Saito}},\ }\bibfield  {title} {\enquote {\bibinfo {title}
  {Observation of dark states in a superconductor diamond quantum hybrid
  system},}\ }\href@noop {} {\bibfield  {journal} {\bibinfo  {journal} {Nature
  Comms.}\ }\textbf {\bibinfo {volume} {3424}},\ \bibinfo {pages} {4524}
  (\bibinfo {year} {2014})}\BibitemShut {NoStop}%
\bibitem [{\citenamefont {Skribanowitz}\ \emph {et~al.}(1973)\citenamefont
  {Skribanowitz}, \citenamefont {Herman}, \citenamefont {MacGillivray},\ and\
  \citenamefont {Feld}}]{skribanowitz1973observation}%
  \BibitemOpen
  \bibfield  {author} {\bibinfo {author} {\bibfnamefont {N.}~\bibnamefont
  {Skribanowitz}}, \bibinfo {author} {\bibfnamefont {I.~P.}\ \bibnamefont
  {Herman}}, \bibinfo {author} {\bibfnamefont {J.~C.}\ \bibnamefont
  {MacGillivray}}, \ and\ \bibinfo {author} {\bibfnamefont {M.~S.}\
  \bibnamefont {Feld}},\ }\bibfield  {title} {\enquote {\bibinfo {title}
  {Observation of {D}icke superradiance in optically pumped {HF} gas},}\
  }\href@noop {} {\bibfield  {journal} {\bibinfo  {journal} {Phys. Rev. Lett.}\
  }\textbf {\bibinfo {volume} {30}},\ \bibinfo {pages} {309} (\bibinfo {year}
  {1973})}\BibitemShut {NoStop}%
\bibitem [{\citenamefont {Gross}\ \emph {et~al.}(1976)\citenamefont {Gross},
  \citenamefont {Fabre}, \citenamefont {Pillet},\ and\ \citenamefont
  {Haroche}}]{gross1976observation}%
  \BibitemOpen
  \bibfield  {author} {\bibinfo {author} {\bibfnamefont {M.}~\bibnamefont
  {Gross}}, \bibinfo {author} {\bibfnamefont {C.}~\bibnamefont {Fabre}},
  \bibinfo {author} {\bibfnamefont {P.}~\bibnamefont {Pillet}}, \ and\ \bibinfo
  {author} {\bibfnamefont {S.}~\bibnamefont {Haroche}},\ }\bibfield  {title}
  {\enquote {\bibinfo {title} {Observation of near-infrared {D}icke
  superradiance on cascading transitions in atomic sodium},}\ }\href@noop {}
  {\bibfield  {journal} {\bibinfo  {journal} {Phys. Rev. Lett.}\ }\textbf
  {\bibinfo {volume} {36}},\ \bibinfo {pages} {1035} (\bibinfo {year}
  {1976})}\BibitemShut {NoStop}%
\bibitem [{\citenamefont {Raimond}\ \emph {et~al.}(1982)\citenamefont
  {Raimond}, \citenamefont {Goy}, \citenamefont {Gross}, \citenamefont
  {Fabre},\ and\ \citenamefont {Haroche}}]{raimond1982collective}%
  \BibitemOpen
  \bibfield  {author} {\bibinfo {author} {\bibfnamefont {J.~M.}\ \bibnamefont
  {Raimond}}, \bibinfo {author} {\bibfnamefont {P.}~\bibnamefont {Goy}},
  \bibinfo {author} {\bibfnamefont {M.}~\bibnamefont {Gross}}, \bibinfo
  {author} {\bibfnamefont {C.}~\bibnamefont {Fabre}}, \ and\ \bibinfo {author}
  {\bibfnamefont {S.}~\bibnamefont {Haroche}},\ }\bibfield  {title} {\enquote
  {\bibinfo {title} {Collective absorption of blackbody radiation by {R}ydberg
  atoms in a cavity: an experiment on {B}ose statistics and {B}rownian
  motion},}\ }\href@noop {} {\bibfield  {journal} {\bibinfo  {journal} {Phys.
  Rev. Lett.}\ }\textbf {\bibinfo {volume} {49}},\ \bibinfo {pages} {117}
  (\bibinfo {year} {1982})}\BibitemShut {NoStop}%
\bibitem [{\citenamefont {Scheibner}\ \emph {et~al.}(2007)\citenamefont
  {Scheibner}, \citenamefont {Schmidt}, \citenamefont {Worschech},
  \citenamefont {Forchel}, \citenamefont {Bacher}, \citenamefont {Passow},\
  and\ \citenamefont {Hommel}}]{scheibner2007superradiance}%
  \BibitemOpen
  \bibfield  {author} {\bibinfo {author} {\bibfnamefont {M.}~\bibnamefont
  {Scheibner}}, \bibinfo {author} {\bibfnamefont {T.}~\bibnamefont {Schmidt}},
  \bibinfo {author} {\bibfnamefont {L.}~\bibnamefont {Worschech}}, \bibinfo
  {author} {\bibfnamefont {A.}~\bibnamefont {Forchel}}, \bibinfo {author}
  {\bibfnamefont {G.}~\bibnamefont {Bacher}}, \bibinfo {author} {\bibfnamefont
  {T.}~\bibnamefont {Passow}}, \ and\ \bibinfo {author} {\bibfnamefont
  {D.}~\bibnamefont {Hommel}},\ }\bibfield  {title} {\enquote {\bibinfo {title}
  {Superradiance of quantum dots},}\ }\href@noop {} {\bibfield  {journal}
  {\bibinfo  {journal} {Nature Physics}\ }\textbf {\bibinfo {volume} {3}},\
  \bibinfo {pages} {106--110} (\bibinfo {year} {2007})}\BibitemShut {NoStop}%
\bibitem [{\citenamefont {R{\"o}hlsberger}\ \emph {et~al.}(2010)\citenamefont
  {R{\"o}hlsberger}, \citenamefont {Schlage}, \citenamefont {Sahoo},
  \citenamefont {Couet},\ and\ \citenamefont
  {R{\"u}ffer}}]{rohlsberger2010collective}%
  \BibitemOpen
  \bibfield  {author} {\bibinfo {author} {\bibfnamefont {Ralf}\ \bibnamefont
  {R{\"o}hlsberger}}, \bibinfo {author} {\bibfnamefont {Kai}\ \bibnamefont
  {Schlage}}, \bibinfo {author} {\bibfnamefont {Balaram}\ \bibnamefont
  {Sahoo}}, \bibinfo {author} {\bibfnamefont {Sebastien}\ \bibnamefont
  {Couet}}, \ and\ \bibinfo {author} {\bibfnamefont {Rudolf}\ \bibnamefont
  {R{\"u}ffer}},\ }\bibfield  {title} {\enquote {\bibinfo {title} {Collective
  lamb shift in single-photon superradiance},}\ }\href@noop {} {\bibfield
  {journal} {\bibinfo  {journal} {Science}\ }\textbf {\bibinfo {volume}
  {328}},\ \bibinfo {pages} {1248--1251} (\bibinfo {year} {2010})}\BibitemShut
  {NoStop}%
\bibitem [{\citenamefont {DeVoe}\ and\ \citenamefont
  {Brewer}(1996)}]{devoe1996observation}%
  \BibitemOpen
  \bibfield  {author} {\bibinfo {author} {\bibfnamefont {R.~G.}\ \bibnamefont
  {DeVoe}}\ and\ \bibinfo {author} {\bibfnamefont {R.~G.}\ \bibnamefont
  {Brewer}},\ }\bibfield  {title} {\enquote {\bibinfo {title} {Observation of
  superradiant and subradiant spontaneous emission of two trapped ions},}\
  }\href@noop {} {\bibfield  {journal} {\bibinfo  {journal} {Phys. Rev. Lett.}\
  }\textbf {\bibinfo {volume} {76}},\ \bibinfo {pages} {2049} (\bibinfo {year}
  {1996})}\BibitemShut {NoStop}%
\bibitem [{\citenamefont {Eschner}\ \emph {et~al.}(2001)\citenamefont
  {Eschner}, \citenamefont {Raab}, \citenamefont {Schmidt-Kaler},\ and\
  \citenamefont {Blatt}}]{eschner2001light}%
  \BibitemOpen
  \bibfield  {author} {\bibinfo {author} {\bibfnamefont {J.}~\bibnamefont
  {Eschner}}, \bibinfo {author} {\bibfnamefont {Ch.}\ \bibnamefont {Raab}},
  \bibinfo {author} {\bibfnamefont {F.}~\bibnamefont {Schmidt-Kaler}}, \ and\
  \bibinfo {author} {\bibfnamefont {R.}~\bibnamefont {Blatt}},\ }\bibfield
  {title} {\enquote {\bibinfo {title} {Light interference from single atoms and
  their mirror images},}\ }\href@noop {} {\bibfield  {journal} {\bibinfo
  {journal} {Nature}\ }\textbf {\bibinfo {volume} {413}},\ \bibinfo {pages}
  {495--498} (\bibinfo {year} {2001})}\BibitemShut {NoStop}%
\bibitem [{\citenamefont {Filipp}\ \emph {et~al.}(2011)\citenamefont {Filipp},
  \citenamefont {G{\"o}ppl}, \citenamefont {Fink}, \citenamefont {Baur},
  \citenamefont {Bianchetti}, \citenamefont {Steffen},\ and\ \citenamefont
  {Wallraff}}]{filipp2011multimode}%
  \BibitemOpen
  \bibfield  {author} {\bibinfo {author} {\bibfnamefont {S.}~\bibnamefont
  {Filipp}}, \bibinfo {author} {\bibfnamefont {M.}~\bibnamefont {G{\"o}ppl}},
  \bibinfo {author} {\bibfnamefont {J.~M.}\ \bibnamefont {Fink}}, \bibinfo
  {author} {\bibfnamefont {M.}~\bibnamefont {Baur}}, \bibinfo {author}
  {\bibfnamefont {R.}~\bibnamefont {Bianchetti}}, \bibinfo {author}
  {\bibfnamefont {L.}~\bibnamefont {Steffen}}, \ and\ \bibinfo {author}
  {\bibfnamefont {A.}~\bibnamefont {Wallraff}},\ }\bibfield  {title} {\enquote
  {\bibinfo {title} {Multimode mediated qubit-qubit coupling and dark-state
  symmetries in circuit quantum electrodynamics},}\ }\href@noop {} {\bibfield
  {journal} {\bibinfo  {journal} {Phys. Rev. A}\ }\textbf {\bibinfo {volume}
  {83}},\ \bibinfo {pages} {063827} (\bibinfo {year} {2011})}\BibitemShut
  {NoStop}%
\bibitem [{\citenamefont {Van~Loo}\ \emph {et~al.}(2013)\citenamefont
  {Van~Loo}, \citenamefont {Fedorov}, \citenamefont {Lalumi{\`e}re},
  \citenamefont {Sanders}, \citenamefont {Blais},\ and\ \citenamefont
  {Wallraff}}]{van2013photon}%
  \BibitemOpen
  \bibfield  {author} {\bibinfo {author} {\bibfnamefont {A.~F.}\ \bibnamefont
  {Van~Loo}}, \bibinfo {author} {\bibfnamefont {A.}~\bibnamefont {Fedorov}},
  \bibinfo {author} {\bibfnamefont {K.}~\bibnamefont {Lalumi{\`e}re}}, \bibinfo
  {author} {\bibfnamefont {B.~C.}\ \bibnamefont {Sanders}}, \bibinfo {author}
  {\bibfnamefont {A.}~\bibnamefont {Blais}}, \ and\ \bibinfo {author}
  {\bibfnamefont {A.}~\bibnamefont {Wallraff}},\ }\bibfield  {title} {\enquote
  {\bibinfo {title} {Photon-mediated interactions between distant artificial
  atoms},}\ }\href@noop {} {\bibfield  {journal} {\bibinfo  {journal}
  {Science}\ }\textbf {\bibinfo {volume} {342}},\ \bibinfo {pages} {1494--1496}
  (\bibinfo {year} {2013})}\BibitemShut {NoStop}%
\bibitem [{\citenamefont {Mlynek}\ \emph {et~al.}(2014)\citenamefont {Mlynek},
  \citenamefont {Abdumalikov}, \citenamefont {Eichler},\ and\ \citenamefont
  {Wallraff}}]{mlynek2014observation}%
  \BibitemOpen
  \bibfield  {author} {\bibinfo {author} {\bibfnamefont {J.~A.}\ \bibnamefont
  {Mlynek}}, \bibinfo {author} {\bibfnamefont {A.~A.}\ \bibnamefont
  {Abdumalikov}}, \bibinfo {author} {\bibfnamefont {C.}~\bibnamefont
  {Eichler}}, \ and\ \bibinfo {author} {\bibfnamefont {A.}~\bibnamefont
  {Wallraff}},\ }\bibfield  {title} {\enquote {\bibinfo {title} {Observation of
  {D}icke superradiance for two artificial atoms in a cavity with high decay
  rate},}\ }\href@noop {} {\bibfield  {journal} {\bibinfo  {journal} {Nature
  Comms.}\ }\textbf {\bibinfo {volume} {5}} (\bibinfo {year}
  {2014})}\BibitemShut {NoStop}%
\bibitem [{\citenamefont {Meiser}\ and\ \citenamefont
  {Holland}(2010)}]{Meiser2010}%
  \BibitemOpen
  \bibfield  {author} {\bibinfo {author} {\bibfnamefont {D.}~\bibnamefont
  {Meiser}}\ and\ \bibinfo {author} {\bibfnamefont {M.~J.}\ \bibnamefont
  {Holland}},\ }\bibfield  {title} {\enquote {\bibinfo {title} {Steady-state
  superradiance with alkaline-earth-metal atoms},}\ }\href {\doibase
  10.1103/PhysRevA.81.033847} {\bibfield  {journal} {\bibinfo  {journal} {Phys.
  Rev. A}\ }\textbf {\bibinfo {volume} {81}},\ \bibinfo {pages} {033847}
  (\bibinfo {year} {2010})}\BibitemShut {NoStop}%
\bibitem [{\citenamefont {Keaveney}\ \emph {et~al.}(2012)\citenamefont
  {Keaveney}, \citenamefont {Sargsyan}, \citenamefont {Krohn}, \citenamefont
  {Hughes}, \citenamefont {Sarkisyan},\ and\ \citenamefont
  {Adams}}]{keaveney2012cooperative}%
  \BibitemOpen
  \bibfield  {author} {\bibinfo {author} {\bibfnamefont {James}\ \bibnamefont
  {Keaveney}}, \bibinfo {author} {\bibfnamefont {Armen}\ \bibnamefont
  {Sargsyan}}, \bibinfo {author} {\bibfnamefont {Ulrich}\ \bibnamefont
  {Krohn}}, \bibinfo {author} {\bibfnamefont {Ifan~G}\ \bibnamefont {Hughes}},
  \bibinfo {author} {\bibfnamefont {David}\ \bibnamefont {Sarkisyan}}, \ and\
  \bibinfo {author} {\bibfnamefont {Charles~S}\ \bibnamefont {Adams}},\
  }\bibfield  {title} {\enquote {\bibinfo {title} {Cooperative lamb shift in an
  atomic vapor layer of nanometer thickness},}\ }\href@noop {} {\bibfield
  {journal} {\bibinfo  {journal} {Physical review letters}\ }\textbf {\bibinfo
  {volume} {108}},\ \bibinfo {pages} {173601} (\bibinfo {year}
  {2012})}\BibitemShut {NoStop}%
\bibitem [{\citenamefont {Roof}\ \emph {et~al.}(2016)\citenamefont {Roof},
  \citenamefont {Kemp}, \citenamefont {Havey},\ and\ \citenamefont
  {Sokolov}}]{roof2016observation}%
  \BibitemOpen
  \bibfield  {author} {\bibinfo {author} {\bibfnamefont {SJ}~\bibnamefont
  {Roof}}, \bibinfo {author} {\bibfnamefont {KJ}~\bibnamefont {Kemp}}, \bibinfo
  {author} {\bibfnamefont {MD}~\bibnamefont {Havey}}, \ and\ \bibinfo {author}
  {\bibfnamefont {IM}~\bibnamefont {Sokolov}},\ }\bibfield  {title} {\enquote
  {\bibinfo {title} {Observation of single-photon superradiance and the
  cooperative lamb shift in an extended sample of cold atoms},}\ }\href@noop {}
  {\bibfield  {journal} {\bibinfo  {journal} {arXiv preprint arXiv:1603.07268}\
  } (\bibinfo {year} {2016})}\BibitemShut {NoStop}%
\bibitem [{\citenamefont {Orlando}\ \emph {et~al.}(1999)\citenamefont
  {Orlando}, \citenamefont {Mooij}, \citenamefont {Tian}, \citenamefont
  {van~der Wal}, \citenamefont {Levitov}, \citenamefont {Lloyd},\ and\
  \citenamefont {Mazo}}]{orlando1999superconducting}%
  \BibitemOpen
  \bibfield  {author} {\bibinfo {author} {\bibfnamefont {T.~P.}\ \bibnamefont
  {Orlando}}, \bibinfo {author} {\bibfnamefont {J.~E.}\ \bibnamefont {Mooij}},
  \bibinfo {author} {\bibfnamefont {L.}~\bibnamefont {Tian}}, \bibinfo {author}
  {\bibfnamefont {C.~H.}\ \bibnamefont {van~der Wal}}, \bibinfo {author}
  {\bibfnamefont {L.~S.}\ \bibnamefont {Levitov}}, \bibinfo {author}
  {\bibfnamefont {S.}~\bibnamefont {Lloyd}}, \ and\ \bibinfo {author}
  {\bibfnamefont {J.~J.}\ \bibnamefont {Mazo}},\ }\bibfield  {title} {\enquote
  {\bibinfo {title} {Superconducting persistent-current qubit},}\ }\href@noop
  {} {\bibfield  {journal} {\bibinfo  {journal} {Phys. Rev. B}\ }\textbf
  {\bibinfo {volume} {60}},\ \bibinfo {pages} {15398} (\bibinfo {year}
  {1999})}\BibitemShut {NoStop}%
\bibitem [{\citenamefont {Johansson}\ \emph {et~al.}(2012)\citenamefont
  {Johansson}, \citenamefont {Nation},\ and\ \citenamefont {Nori}}]{qutip1}%
  \BibitemOpen
  \bibfield  {author} {\bibinfo {author} {\bibfnamefont {J.~R.}\ \bibnamefont
  {Johansson}}, \bibinfo {author} {\bibfnamefont {P.~D.}\ \bibnamefont
  {Nation}}, \ and\ \bibinfo {author} {\bibfnamefont {F.}~\bibnamefont
  {Nori}},\ }\bibfield  {title} {\enquote {\bibinfo {title} {{QuTiP: An
  open-source Python framework for the dynamics of open quantum systems}},}\
  }\href {\doibase 10.1016/j.cpc.2012.02.021} {\bibfield  {journal} {\bibinfo
  {journal} {Computer Physics Communications}\ }\textbf {\bibinfo {volume}
  {183}},\ \bibinfo {pages} {1760} (\bibinfo {year} {2012})},\ \Eprint
  {http://arxiv.org/abs/1110.0573} {1110.0573} \BibitemShut {NoStop}%
\bibitem [{\citenamefont {Johansson}\ \emph {et~al.}(2013)\citenamefont
  {Johansson}, \citenamefont {Nation},\ and\ \citenamefont {Nori}}]{qutip2}%
  \BibitemOpen
  \bibfield  {author} {\bibinfo {author} {\bibfnamefont {J.~R.}\ \bibnamefont
  {Johansson}}, \bibinfo {author} {\bibfnamefont {P.~D.}\ \bibnamefont
  {Nation}}, \ and\ \bibinfo {author} {\bibfnamefont {F.}~\bibnamefont
  {Nori}},\ }\bibfield  {title} {\enquote {\bibinfo {title} {{QuTiP 2: A Python
  framework for the dynamics of open quantum systems}},}\ }\href {\doibase
  10.1016/j.cpc.2012.11.019} {\bibfield  {journal} {\bibinfo  {journal}
  {Computer Physics Communications}\ }\textbf {\bibinfo {volume} {184}},\
  \bibinfo {pages} {1234} (\bibinfo {year} {2013})},\ \Eprint
  {http://arxiv.org/abs/1211.6518} {1211.6518} \BibitemShut {NoStop}%
\bibitem [{\citenamefont {Temnov}\ and\ \citenamefont
  {Woggon}(2005)}]{Temnov05}%
  \BibitemOpen
  \bibfield  {author} {\bibinfo {author} {\bibfnamefont {V.~V.}\ \bibnamefont
  {Temnov}}\ and\ \bibinfo {author} {\bibfnamefont {U.}~\bibnamefont
  {Woggon}},\ }\bibfield  {title} {\enquote {\bibinfo {title} {Superradiance
  and subradiance in an inhomogeneously broadened ensemble of two-level systems
  coupled to a low-$q$ cavity},}\ }\href {\doibase
  10.1103/PhysRevLett.95.243602} {\bibfield  {journal} {\bibinfo  {journal}
  {Phys. Rev. Lett.}\ }\textbf {\bibinfo {volume} {95}},\ \bibinfo {pages}
  {243602} (\bibinfo {year} {2005})}\BibitemShut {NoStop}%
\bibitem [{\citenamefont {Delanty}\ \emph {et~al.}(2011)\citenamefont
  {Delanty}, \citenamefont {Rebic},\ and\ \citenamefont {Twamley}}]{Delanty11}%
  \BibitemOpen
  \bibfield  {author} {\bibinfo {author} {\bibfnamefont {M.}~\bibnamefont
  {Delanty}}, \bibinfo {author} {\bibfnamefont {S.}~\bibnamefont {Rebic}}, \
  and\ \bibinfo {author} {\bibfnamefont {J.}~\bibnamefont {Twamley}},\
  }\bibfield  {title} {\enquote {\bibinfo {title} {Superradiance and phase
  multistability in circuit quantum electrodynamics},}\ }\href@noop {}
  {\bibfield  {journal} {\bibinfo  {journal} {New Journal of Physics}\ }\textbf
  {\bibinfo {volume} {13}},\ \bibinfo {pages} {053032} (\bibinfo {year}
  {2011})}\BibitemShut {NoStop}%
\bibitem [{\citenamefont {Bartels}\ and\ \citenamefont
  {Mintert}(2013)}]{mintert2}%
  \BibitemOpen
  \bibfield  {author} {\bibinfo {author} {\bibfnamefont {B.}~\bibnamefont
  {Bartels}}\ and\ \bibinfo {author} {\bibfnamefont {F.}~\bibnamefont
  {Mintert}},\ }\bibfield  {title} {\enquote {\bibinfo {title} {Smooth optimal
  control with {F}loquet theory},}\ }\href {\doibase
  10.1103/PhysRevA.88.052315} {\bibfield  {journal} {\bibinfo  {journal} {Phys.
  Rev. A}\ }\textbf {\bibinfo {volume} {88}},\ \bibinfo {pages} {052315}
  (\bibinfo {year} {2013})}\BibitemShut {NoStop}%
\bibitem [{\citenamefont {N\"obauer}\ \emph {et~al.}(2015)\citenamefont
  {N\"obauer}, \citenamefont {Angerer}, \citenamefont {Bartels}, \citenamefont
  {Trupke}, \citenamefont {Rotter}, \citenamefont {Schmiedmayer}, \citenamefont
  {Mintert},\ and\ \citenamefont {Majer}}]{mintert1}%
  \BibitemOpen
  \bibfield  {author} {\bibinfo {author} {\bibfnamefont {T.}~\bibnamefont
  {N\"obauer}}, \bibinfo {author} {\bibfnamefont {A.}~\bibnamefont {Angerer}},
  \bibinfo {author} {\bibfnamefont {B.}~\bibnamefont {Bartels}}, \bibinfo
  {author} {\bibfnamefont {M.}~\bibnamefont {Trupke}}, \bibinfo {author}
  {\bibfnamefont {S.}~\bibnamefont {Rotter}}, \bibinfo {author} {\bibfnamefont
  {J.}~\bibnamefont {Schmiedmayer}}, \bibinfo {author} {\bibfnamefont
  {F.}~\bibnamefont {Mintert}}, \ and\ \bibinfo {author} {\bibfnamefont
  {J.}~\bibnamefont {Majer}},\ }\bibfield  {title} {\enquote {\bibinfo {title}
  {Smooth optimal quantum control for robust solid-state spin magnetometry},}\
  }\href {\doibase 10.1103/PhysRevLett.115.190801} {\bibfield  {journal}
  {\bibinfo  {journal} {Phys. Rev. Lett.}\ }\textbf {\bibinfo {volume} {115}},\
  \bibinfo {pages} {190801} (\bibinfo {year} {2015})}\BibitemShut {NoStop}%
\bibitem [{\citenamefont {Yoshihara}\ \emph {et~al.}(2014)\citenamefont
  {Yoshihara}, \citenamefont {Nakamura}, \citenamefont {Yan}, \citenamefont
  {Gustavsson}, \citenamefont {Bylander}, \citenamefont {Oliver},\ and\
  \citenamefont {Tsai}}]{yoshihara2014flux}%
  \BibitemOpen
  \bibfield  {author} {\bibinfo {author} {\bibfnamefont {F.}~\bibnamefont
  {Yoshihara}}, \bibinfo {author} {\bibfnamefont {Y.}~\bibnamefont {Nakamura}},
  \bibinfo {author} {\bibfnamefont {F.}~\bibnamefont {Yan}}, \bibinfo {author}
  {\bibfnamefont {S.}~\bibnamefont {Gustavsson}}, \bibinfo {author}
  {\bibfnamefont {J.}~\bibnamefont {Bylander}}, \bibinfo {author}
  {\bibfnamefont {W.~D.}\ \bibnamefont {Oliver}}, \ and\ \bibinfo {author}
  {\bibfnamefont {J.~S.}\ \bibnamefont {Tsai}},\ }\bibfield  {title} {\enquote
  {\bibinfo {title} {Flux qubit noise spectroscopy using {R}abi oscillations
  under strong driving conditions},}\ }\href@noop {} {\bibfield  {journal}
  {\bibinfo  {journal} {Physical Review B}\ }\textbf {\bibinfo {volume} {89}},\
  \bibinfo {pages} {020503} (\bibinfo {year} {2014})}\BibitemShut {NoStop}%
\bibitem [{\citenamefont {Knee}\ \emph {et~al.}(2016)\citenamefont {Knee},
  \citenamefont {Kakuyanagi}, \citenamefont {Yeh}, \citenamefont {Matsuzaki},
  \citenamefont {Toida}, \citenamefont {Yamaguchi}, \citenamefont {Leggett},\
  and\ \citenamefont {Munro}}]{Knee2016}%
  \BibitemOpen
  \bibfield  {author} {\bibinfo {author} {\bibfnamefont {G.~C.}\ \bibnamefont
  {Knee}}, \bibinfo {author} {\bibfnamefont {K.}~\bibnamefont {Kakuyanagi}},
  \bibinfo {author} {\bibfnamefont {M-C.}\ \bibnamefont {Yeh}}, \bibinfo
  {author} {\bibfnamefont {Y.}~\bibnamefont {Matsuzaki}}, \bibinfo {author}
  {\bibfnamefont {H.}~\bibnamefont {Toida}}, \bibinfo {author} {\bibfnamefont
  {H.}~\bibnamefont {Yamaguchi}}, \bibinfo {author} {\bibfnamefont {A.~J.}\
  \bibnamefont {Leggett}}, \ and\ \bibinfo {author} {\bibfnamefont {W.~J.}\
  \bibnamefont {Munro}},\ }\bibfield  {title} {\enquote {\bibinfo {title} {A
  strict experimental test of macroscopic realism in a superconducting flux
  qubit},}\ }\href {http://arxiv.org/abs/1601.03728} {\bibfield  {journal}
  {\bibinfo  {journal} {arXiv: 1601.03728}\ } (\bibinfo {year}
  {2016})}\BibitemShut {NoStop}%
\bibitem [{\citenamefont {Lambert}\ \emph {et~al.}(2016)\citenamefont
  {Lambert}, \citenamefont {Debnath}, \citenamefont {Kockum}, \citenamefont
  {Knee}, \citenamefont {Munro},\ and\ \citenamefont {Nori}}]{lambert16}%
  \BibitemOpen
  \bibfield  {author} {\bibinfo {author} {\bibfnamefont {N.}~\bibnamefont
  {Lambert}}, \bibinfo {author} {\bibfnamefont {K.}~\bibnamefont {Debnath}},
  \bibinfo {author} {\bibfnamefont {A.~Frisk}\ \bibnamefont {Kockum}}, \bibinfo
  {author} {\bibfnamefont {G.~C.}\ \bibnamefont {Knee}}, \bibinfo {author}
  {\bibfnamefont {W.~J.}\ \bibnamefont {Munro}}, \ and\ \bibinfo {author}
  {\bibfnamefont {F.}~\bibnamefont {Nori}},\ }\bibfield  {title} {\enquote
  {\bibinfo {title} {Leggett--{G}arg inequality violations with a large
  ensemble of qubits},}\ }\href@noop {} {\bibfield  {journal} {\bibinfo
  {journal} {Phys. Rev. A}\ }\textbf {\bibinfo {volume} {94}},\ \bibinfo
  {pages} {012105} (\bibinfo {year} {2016})}\BibitemShut {NoStop}%
\end{thebibliography}

%

\end{document}